\documentclass[aps,jmp,amsmath,amssymb,onecolumn,12pt,superscriptaddress]{revtex4-2}

\usepackage{graphicx}
\graphicspath{{FIGURES/}}

\usepackage{dcolumn}% Align TABLE columns on decimal point
\usepackage{bm}% bold math
\usepackage[utf8x]{inputenc}
\usepackage{lipsum}
\usepackage{resizegather}
\usepackage{amssymb}
\usepackage{amsmath}
\usepackage{booktabs}
\usepackage{xcolor}
\usepackage{lineno}

\newcommand{\appropto}{\mathrel{\vcenter{
  \offinterlineskip\halign{\hfil$##$\cr
    \propto\cr\noalign{\kern2pt}\sim\cr\noalign{\kern-2pt}}}}}

\begin{document}
%\date{\today}

\title{Structure and elasticity of model disordered, polydisperse and defect-free polymer networks}

%%% Authors
\author{Valerio Sorichetti}
\affiliation{Laboratoire Charles Coulomb (L2C), Univ. Montpellier, CNRS, F-34095, Montpellier, France}
\affiliation{ IATE, Université de Montpellier, INRAE, Institut Agro, Montpellier, France}
\affiliation{Institute of Science and Technology Austria, 3400 Klosterneuburg, Austria}

\author{Andrea Ninarello}
\affiliation{CNR-ISC Uos Sapienza, Piazzale A. Moro 2, IT-00185 Roma, Italy}
\affiliation{Department of Physics, {\textit Sapienza} Universit\`a di Roma, Piazzale A. Moro 2, IT-00185 Roma, Italy}

\author{Jos\'e Ruiz-Franco}
\affiliation{CNR-ISC Uos Sapienza, Piazzale A. Moro 2, IT-00185 Roma, Italy}
\affiliation{Department of Physics, {\textit Sapienza} Universit\`a di Roma, Piazzale A. Moro 2, IT-00185 Roma, Italy}
\affiliation{Physical Chemistry and Soft Matter, Wageningen University \& Research, Stippeneng 4, 6708WE Wageningen, Netherlands}

\author{Virginie Hugouvieux}
\affiliation{IATE, Université de Montpellier, INRAE, Institut Agro, Montpellier, France}

\author{Emanuela Zaccarelli}
\affiliation{CNR-ISC Uos Sapienza, Piazzale A. Moro 2, IT-00185 Roma, Italy}
\affiliation{Department of Physics, {\textit Sapienza} Universit\`a di Roma, Piazzale A. Moro 2, IT-00185 Roma, Italy}

\author{Cristian Micheletti}
\affiliation{SISSA - Scuola Internazionale Superiore di Studi Avanzati, Via Bonomea 265, 34136 Trieste, Italy}

\author{Walter Kob}
\affiliation{Laboratoire Charles Coulomb (L2C), Univ. Montpellier, CNRS, F-34095, Montpellier, France}
\affiliation{Institut Universitaire de France}

\author{Lorenzo Rovigatti}
\email[Corresponding author: ]{lorenzo.rovigatti@uniroma1.it}
\affiliation{CNR-ISC Uos Sapienza, Piazzale A. Moro 2, IT-00185 Roma, Italy}
\affiliation{Department of Physics, {\textit Sapienza} Universit\`a di Roma, Piazzale A. Moro 2, IT-00185 Roma, Italy}

%%%%%%%%%%%%%%%%%%%%%%%%%%%%%%%%%%%%%%%%%%%%%%%%%%%%%%
\begin{abstract}
The elasticity of disordered and polydisperse polymer networks is a fundamental problem of soft matter physics that is still open.  Here, we self-assemble polymer networks \textit{via} simulations of a mixture of bivalent and tri- or tetravalent patchy particles, which result in an exponential strand length distribution analogous to that of experimental randomly crosslinked systems. After assembly, the network connectivity and topology are frozen and the resulting system is characterized. We find that the fractal structure of the network depends on the number density at which the assembly has been carried out, but that systems with the same mean valence and same assembly density have the same structural properties. Moreover, we compute the long-time limit of the mean-squared displacement, also known as the (squared) localization length, of the crosslinks and of the middle monomers of the strands, showing that the dynamics of long strands is well described by the tube model. Finally, we find a relation connecting these two localization lengths at high density, and connect the crosslink localization length to the shear modulus of the system.

\end{abstract}
%%%%%%%%%%%%%%%%%%%%%%%%%%%%%%%%%%%%%%%%%%%%%%%%%%%%%%
%\pagewiselinenumbers
\maketitle

%%%%%%%%%%%%%%%%%%%%%%%%%%%%%%%%%%%%%%%%%%%%%%%%%%%%%%
%%%%%%%%%%%%%%%%%%%%%%%%%%%%%%%%%%%%%%%%%%%%%%%%%%%%%%
\section{Introduction}
\label{sec:intro}
%%%%%%%%%%%%%%%%%%%%%%%%%%%%%%%%%%%%%%%%%%%%%%%%%%%%%%
%%%%%%%%%%%%%%%%%%%%%%%%%%%%%%%%%%%%%%%%%%%%%%%%%%%%%%

Polymer networks are solids that can be obtained by crosslinking polymeric chains \cite{rubinstein2003polymer}. From rubbers \cite{shanks2013general} to hydrogels \cite{richbourg2020swollen} to biological networks \cite{wen2011polymer,licup2015stress}, these systems have countless industrial \cite{cordier2008self} and biomedical \cite{hoffman2012hydrogels} applications. Understanding how the macroscopic properties of polymer networks, and in particular their elasticity, depend on their structure, chemistry and topology is still a largely open problem \cite{zhong2016quantifying,hoshino2018network,matsuda2019fabrication,lin2019revisiting,gula2020computational,lang2022reference}. To unravel these questions, numerical simulations \cite{grest1992kinetics,duering1994structure,kenkare1998discontinuous,putz2000self,gilra2000monte,lang2017relation,lang2020analysis,lang2022reference,grest1990statistical,duering1991relaxation,duering1992structural,lang2013monomer,gula2020computational,zheng2022understanding,sonnenburg1990molecular,everaers1995test,escobedo1996monte,everaers1996topological,escobedo1997simulation, everaers1999entanglement,sorichetti2021effect,tauber2021sharing,ninarello2022onset} are an invaluable tool, as they allow a control on the structure and topology of the network that is impossible to achieve in experiments, making it possible to disentangle the effects of different microscopic contributions.

Since crosslinked polymer networks are systems with quenched (frozen-in) disorder, their properties depend on the way the network is formed. Therefore, when simulating a model polymer network, the first step is to choose an assembly protocol. One possibility is starting with a system of precursor polymers, which can be mono- or poly-disperse, linear or branched, which are then crosslinked \textit{via} some procedure. The crosslinking can be allowed to occur only between chain ends  (\emph{end-linking}) \cite{grest1992kinetics,duering1994structure,kenkare1998discontinuous,putz2000self,gilra2000monte,lang2017relation,lang2020analysis,lang2022reference,zheng2022understanding} or between any pair of monomers (\emph{random crosslinking}) \cite{grest1990statistical,duering1991relaxation,duering1992structural,lang2013monomer,gula2020computational}. Random crosslinking, however, is only efficient at melt densities, whereas end-linking suffers from kinetic limitations, as reaching a perfect (fully-bonded) configuration requires a time that grows quickly as the length of chains grows \cite{grest1992kinetics,duering1994structure,kenkare1998discontinuous}. \emph{Ad hoc} methods can be used to increase the number of bonded sites \cite{duering1994structure, kenkare1998discontinuous}. Using these methods is however not completely satisfactory, since the final structure in principle depends on the exact method that was used to force the formation of the bonds. Another option is to impose some lattice connectivity, like the diamond lattice \cite{sonnenburg1990molecular,everaers1995test,escobedo1996monte,everaers1996topological,escobedo1997simulation, everaers1999entanglement}. Several of these ``lattice networks" can then be randomly superimposed in order to obtain a disordered structure \cite{everaers1995test,everaers1996topological,everaers1999entanglement}. These systems, however, present an underlying ordered topology and monodisperse strand length, contrary to most experimental systems. 

Here we study a model of disordered, polydisperse and defect-free (\textit{i.e.}, fully bonded) networks, which has originally been developed for the study of microgels \cite{gnan2017silico,rovigatti2017internal,camerin2018modelling,rovigatti2019numerical,rovigatti2019connecting, ninarello2019modeling} and later applied to the study of phantom~\cite{sorichetti2021effect}, double~\cite{tauber2021sharing} and hyper-auxetic networks~\cite{ninarello2022onset}. 
This model has been previously validated against experiments of microgels~\cite{gnan2017silico,ninarello2019modeling} and we therefore consider it as a viable computational counterpart of experimentally realizable hydrogel networks.
The macroscopic properties of these networks, which are self-assembled \textit{via} equilibrium simulations, depend only on a very small number of parameters. Moreover, the self-assembly procedure naturally gives rise to a disordered topology with a well-defined exponential chain length distribution, similar to that of randomly crosslinked networks \cite{grest1990statistical,higgs1988polydisperse} or resulting from step-growth polymerization \cite{zhang2022dispersity}. Here, we use molecular dynamics simulations to characterize the structure and elasticity of these systems, and show how these properties are related to each other.

In Sec.~\ref{sec:model}, we give a detailed description of the simulation model and of the self-assembly procedure. In Sec.~\ref{sec:structure}, we study the structural properties of these networks, such as strand length distribution, radius of gyration of the chains, bond angle distribution and structure factor. In Sec.~\ref{sec:elasticity}, we analyze their elastic properties, connecting them to static observables. We conclude in Sec.~\ref{sec:conclusions} with a discussion of our results and of future perspectives.

%%%%%%%%%%%%%%%%%%%%%%%%%%%%%%%%%%%%%%%%%%%%%%%%%%%%%%
%%%%%%%%%%%%%%%%%%%%%%%%%%%%%%%%%%%%%%%%%%%%%%%%%%%%%%
\section{Model and methods}
\label{sec:model}
%%%%%%%%%%%%%%%%%%%%%%%%%%%%%%%%%%%%%%%%%%%%%%%%%%%%%%
%%%%%%%%%%%%%%%%%%%%%%%%%%%%%%%%%%%%%%%%%%%%%%%%%%%%%%

We generate a polydisperse network \emph{via} the method described in Ref. \cite{gnan2017silico}. This method was originally developed for MD simulations of microgels \cite{gnan2017silico,rovigatti2017internal,camerin2018modelling,rovigatti2019numerical,rovigatti2019connecting, ninarello2019modeling}, but can be generalized to the case of bulk systems \cite{sorichetti2021effect,tauber2021sharing,ninarello2022onset}. In contrast to network-generation approaches using direct simulations of cross-linking dynamics \cite{grest1992kinetics,duering1994structure,kenkare1998discontinuous,putz2000self,gilra2000monte,lang2017relation,lang2020analysis,lang2022reference,zheng2022understanding,grest1990statistical,duering1991relaxation,duering1992structural,lang2013monomer,gula2020computational} or based on compenetrating lattices \cite{everaers1995test,everaers1996topological,everaers1999entanglement}, this method allows for an efficient generation of fully-bonded networks by using a bottom-up self-assembly approach~\cite{gao1995efficient} based on a bond-swapping potential, as detailed below.

%%%%%%%%%%%%%%%%%%%%%%%%%%%%%%%%%%%%%%%%%%%%%%%%%%%%%%
\subsection{Network assembly using patchy particles}
%%%%%%%%%%%%%%%%%%%%%%%%%%%%%%%%%%%%%%%%%%%%%%%%%%%%%%

The starting point is a mixture of two different species of \emph{patchy particles}, \textit{i.e.}, spheres of identical size and mass decorated by a certain number of interaction sites (the ``patches") arranged in a regular configuration. The number of patches per particle is called the \emph{valence}. We consider systems of volume $V_\text{\rm init}$ containing $M_\text{tot}=M_2+M_f$ particles, or \textit{monomers}, with $M_2$ bivalent particles and $M_f$ $f$-valent particles. Two patchy particles are reversibly attached to each other when at least two of their respective patches are overlapping. Only pairs formed by two bivalent particles or by a bivalent particle and an $f$-valent particle can be attached to each other, so that crosslinks can be connected only through chains made of bivalent particles to form branched structures. In the following we consider the cases $f = 3$ and $f = 4$.

The interaction potential between a pair of particles $i$ and $j$ is 

\begin{equation}
\mathcal U (\mathbf r_{ij}, \{\mathbf p_i\}, \{\mathbf p_j\}) = \mathcal U_\text{WCA} (r_{ij}) + \sum_{\mathbf{R}_\mu \in \{\mathbf p_i\}} \sum_{\mathbf{R}_\nu \in \{\mathbf p_j\}} \mathcal U_\text{patch} (R_{\mu \nu}),
\label{eq:patchy_pot}
\end{equation}

\noindent where $\mathbf r_{ij} = \mathbf r_i - \mathbf r_j$ is the particle-particle distance, $r_{ij} \equiv |\mathbf r_{ij}|$, $ \{\mathbf p_i\}$, $ \{\mathbf p_j\}$ are the sets of unit vectors identifying the patches of particles $i$ and $j$, respectively, $\mathbf R_{\mu}$ and $\mathbf R_{\nu}$ are the positions of patch $\mu$ on particle $i$ and patch $\nu$ on particle $j$ and $R_{\mu \nu} = |\mathbf r_{ij} + \mathbf R_{\mu} - \mathbf R_{\nu}|$ is their distance. The potential $ \mathcal U_\text{WCA}$ is a Lennard-Jones potential cut and shifted at the minimum to be purely repulsive \cite{weeks1971role}:

\begin{equation}
\mathcal U_{WCA}(r_{ij}) = 
\begin{cases}
4 \epsilon \left[ \left(\frac \sigma {r_{ij}}\right)^{12} -\left(\frac \sigma {r_{ij}}\right)^6+ \frac 1 4 \right] & r_{ij} \leq 2^{1/6} \sigma\\
0 & \text{otherwise}.\\
\end{cases}
\label{eq:wca}
\end{equation}

\begin{figure}
\centering
\includegraphics[width=0.8 \textwidth]{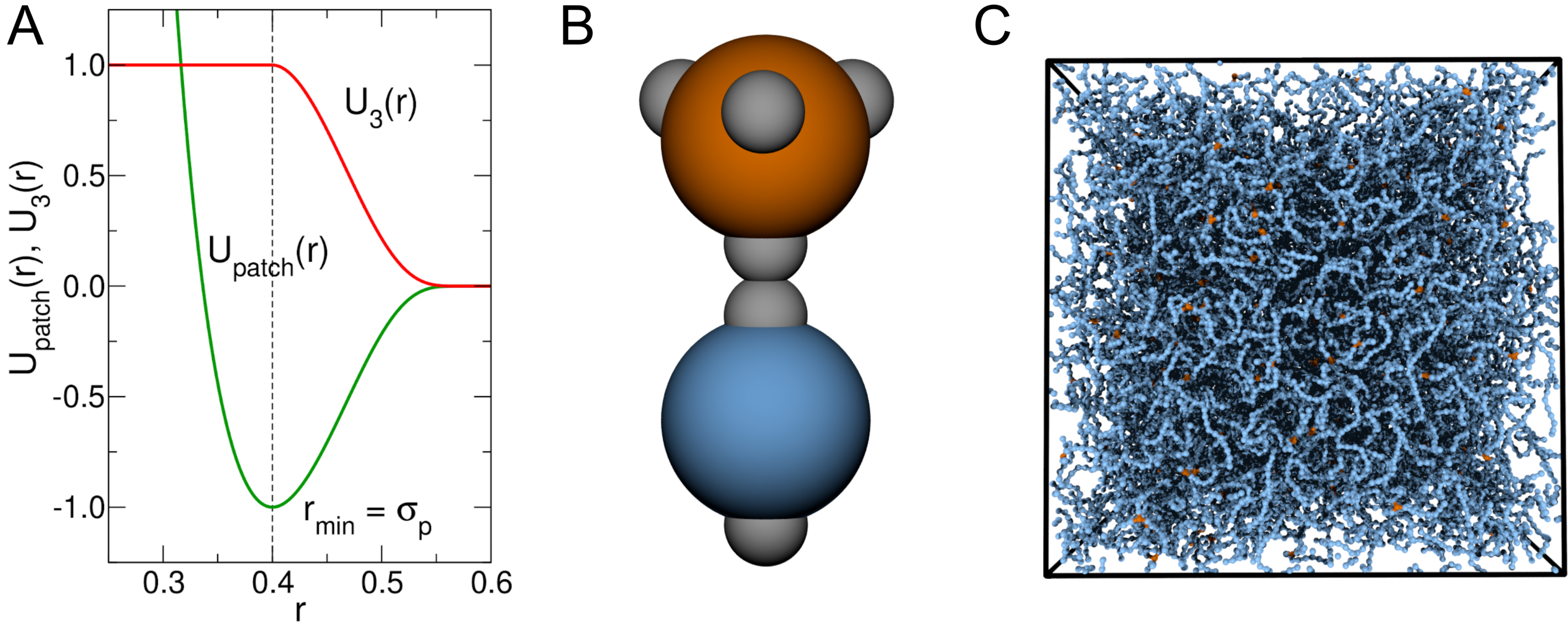}
\caption{(A) Patch potential, Eq.~\eqref{eq:patch_potential}, and $\mathcal U_3$ potential, Eq.~\eqref{eq:pot_u3}. (B) Schematics of a tetravalent ($f=4$, orange) patchy particle attached to a bivalent ($f=2$, blue) particle. The grey spheres represent the patches. (C) Snapshot of one of the simulated networks ($f=3$, $c=1\%$, $\rho_{\rm init}=0.1$, $\rho = 0.97$). Bivalent particles are shown in blue, while crosslinks are shown in orange. Note that some of the particles look detached from the network because of periodic-boundary conditions.}
\label{fig:snapshots}
\end{figure}

\noindent
In the following, all quantities are given in reduced units, with the units of energy, length and mass are thus, respectively, $\epsilon$, $\sigma$ and $m$, where $m$ is the mass of a monomer. The units of temperature and time are, respectively, $T^*=\epsilon/k_B$ and $\tau^*=\sqrt{m \sigma^2/\epsilon}$, where $k_B$ is Boltzmann's constant, which we set equal to $1$. The patch potential takes the form

\begin{equation}
\mathcal U_\text{patch} (R_{\mu \nu}) = 
\begin{cases}
2 \epsilon_{\mu \nu} \left( \frac{\sigma_p^4}{2 R_{\mu \nu}^4} -1 \right) \exp\left(\frac{\sigma_p}{R_{\mu \nu}-r_c} + 2\right)& R_{\mu \nu}<r_c\\
0 & \text{otherwise},\\
\end{cases}
\label{eq:patch_potential}
\end{equation}

\noindent where $r_c=1.5\sigma_p$ (and therefore $r_c$ is the distance at which $\mathcal U_\text{patch}(r_c) = 0$) and $\sigma_p$ is the position of the minimum of the attractive well of depth $\epsilon_{\mu \nu}$, which we set to $\sigma_p = 0.4$. The resulting interaction potential is shown in Fig.~\ref{fig:snapshots}A. The interaction energy $\epsilon_{\mu \nu}$ is $\epsilon_{\mu \nu}=1$ for all pairs of patches, except for pairs of $f$-valent particles, for which $\epsilon_{\mu \nu}=0$, so that the bonding between two $f$-valent particles (crosslinks) is forbidden. The patches are arranged on the poles, equidistant on the equator, and on a tetrahedron for bi-, tri- and tetravalent particles, respectively (see for instance Fig.~\ref{fig:snapshots}B). In all cases, the distance between the patch and the center of the particle is $1/2$. The pair potential given in Eq.~\eqref{eq:patch_potential} is complemented by a three-body potential $\mathcal U_\text{triplet}$ acting on triplets of nearby patches \cite{sciortino2017three}:

\begin{equation}
\mathcal U_\text{triplet} = w \sum_{\lambda,\mu,\nu} \epsilon_{\mu \nu} \mathcal U_3 (r_{\lambda,\mu}) \mathcal U_3 (r_{\lambda,\nu}),
\label{eq:pot_triplet}
\end{equation}

\noindent where $\mathcal U_3 (r)$ has the following form, also shown in Fig.~\ref{fig:snapshots}A:

\begin{equation}
\mathcal U_3 (r) = 
\begin{cases}
1 & r < r_\text{min}\\
- \mathcal U_\text{patch} (r) / \epsilon_{\mu \nu} & r_\text{min} < r < r_c.
\end{cases}
\label{eq:pot_u3}
\end{equation}

\noindent The term \eqref{eq:pot_triplet} has a twofold effect: On one hand it enforces the single-bond-per-patch condition: a given patch cannot be involved in more than one bond at a time. On the other hand, the three-body term also allows to introduce an efficient bond-swapping mechanism that makes it possible to easily equilibrate the system at extremely low temperatures. The parameter $w$ appearing in Eq.~\eqref{eq:pot_triplet} can be used to tune the amplitude of $\mathcal U_\text{triplet}$, in order to favor ($w \simeq 1$) or hamper ($w \gg 1$) bond swapping \cite{sciortino2017three}. 

The assembly is performed \textit{via} molecular dynamics simulations (GPU implementation of the oxDNA software \cite{rovigatti2015comparison}) run in the $NVT$ ensemble at $T=T_\text{assembly}=0.05$. At this low temperature, the system approaches the fully-bonded ground state. Since the bonds can break and reform with an efficiency that is greatly improved by the bond-swapping potential, Eq.~\eqref{eq:pot_triplet}, the system can quickly reach equilibrium. Thus, for a given $T_\text{assembly}$, the properties of the final state are uniquely determined by only three parameters, as it will be shown below: the fraction of crosslinks $c=M_f/M_\text{tot}$, the crosslink valence $f$, and the number density $\rho_{\rm init}=M_\text{tot}/V_\text{\rm init}$ at which the assembly has been carried out. We note that the variation of the latter parameter can be regarded as a crude way for tuning solvent quality. In fact, from a mean-field-like perspective an increase in the assembly density corresponds to a stronger effective monomer-monomer interaction.

Once the majority of the bonds ($>99.8\%$) are formed, the simulation is stopped and the particles which do not belong to the percolating cluster (at most $4\%$ of the total in all the simulated systems) are removed. Although we chose for practical reasons to stop the reaction before reaching the fully-bonded ground state of the system, reaching this state is in principle possible with a greater computational effort. 

The so-obtained networks still contain a few dangling ends. As only a very small fraction of particles is part of these loose strands, we do not expect them to have a significant effect on the elastic properties of the network. However, due to their very long relaxation times, which are known to grow exponentially with the length of the strands~\cite{curro1983theoretical,doi1988theory}, they could lead to a slow relaxation of the monomer mean-squared displacement, which is analyzed below. To avoid these long relaxation times, we decide to remove all the (simple and branched) dangling ends in the system, obtaining a perfect fully-bonded network. We note that for systems with tetravalent crosslinks the removal of the dangling ends has the side effect of introducing some trivalent crosslinks in the system, since when a dangling end is cut a crosslink loses a bond. In this case, the crosslink fraction should be calculated as $c=(M_3+M_4)/(M_2+M_3+M_4)$. However, for all the systems considered, with the exception of $c = 1\%$, where the number of dangling ends can become comparable to that of crosslinkers, we have $M_3 \ll M_4$ and therefore the presence of the trivalent crosslinks is negligible. By contrast, in the case of trivalent crosslinks the removal of dangling ends transforms crosslinks into regular bifunctional particles. Regardless, in all cases at the end of the procedure we obtain a network which is naturally disordered and almost fully-bonded \cite{sorichetti2021effect,ninarello2022onset}. 

%%%%%%%%%%%%%%%%%%%%%%%%%%%%%%%%%%%%%%%%%%%%%%%%%%%%%%
\subsection{$NPT$ simulations with frozen topology and connectivity}
%%%%%%%%%%%%%%%%%%%%%%%%%%%%%%%%%%%%%%%%%%%%%%%%%%%%%%

Once the dangling ends are removed, the interaction potential of Eq.~\eqref{eq:patchy_pot} is replaced by the Kremer-Grest potential \cite{bishop1979molecular,kremer1990dynamics}: The excluded volume interaction is still given by the WCA potential, Eq.~\eqref{eq:wca}, but the reversible bonds of the patchy system are replaced by permanent FENE bonds, with interaction potential

\begin{equation}
\mathcal U_\text{FENE}(r_{ij})= -\frac {k r_0^2} 2 \ln \left[1-\left(r_{ij}/{r_0}\right)^2\right],
\end{equation}

\noindent where $k=30 \epsilon/\sigma^2$ and $r_0 = 1.5 \sigma$. The combined effect of the FENE and the WCA potentials prevents chain crossing at the thermodynamic conditions considered here, so that both the topology and connectivity of the system are frozen~\cite{kremer1990dynamics}.

We consider systems containing an initial number of particles (before the assembly of the network) $M_\text{tot}=5\cdot10^4$, with initial crosslink fractions $c=1\%,5\%$, and $10\%$, and two different crosslink valences, $f=3$ (trivalent) and $f=4$. For $c=5\%$ and $10\%$, we build the network starting from initial number densities $\rho_\text{\rm init}=0.05, 0.1, 0.2, 0.3, 0.4, 0.5$, and $0.85$ (the latter is chosen to mimic typical melt concentrations \cite{kremer1990dynamics}, and thus an elastomer-like system), whereas for $c=1\%$ we only considered $\rho_\text{\rm init}=0.05,0.1$ and $0.2$, as the network phase separates at higher densities (see discussion below). For each system, we consider two independent realizations to make sure that their properties do not depend in a relevant manner from the initial conditions. Moreover, for $c=5\%,10\%$ and $\rho_\text{init}=0.1,0.2$ we have also considered systems of $4\cdot10^5$ particles in order to check for the presence of significant finite size effects, which were found to be absent for the structural and dynamic quantities considered in this work. We note that not all combinations of $f$, $c$ and $\rho_\text{init}$ will generate a homogeneous percolating network. This is because the thermodynamics of systems of limited-valence particles like those used here to generate the networks is controlled by the mean valence \cite{wertheim1984fluids1,wertheim1984fluids2,zaccarelli2005model,bianchi2006phase,sciortino2011reversible, rovigatti2013computing}, defined as \footnote{In the systems with tetravalent crosslinks in which some trivalent crosslinks are produced after the removal of the free ends, we take the presence of trivalent crosslinks into account when calculating the mean valence.}

\begin{equation}
F \equiv \frac{2 M_2 + f M_f}{M_2 + M_f}  = 2 + (f-2) c.
\label{eq:mean_valence}
\end{equation}

\noindent
The larger the value of $F$, the larger the density at which the system undergoes a liquid-gas phase separation. Moreover, upon approaching the phase separation boundary, the self-assembled networks becomes increasingly heterogeneous and thus we make sure to be far enough from this boundary by looking at structural quantities such as the structure factor (see below). The compositions and densities of all the systems studied are reported in detail in the Appendix (Tables~\ref{tab:f3} and ~\ref{tab:f4}). 

Molecular dynamics simulations of the network with frozen topology and connectivity are run using the LAMMPS package \cite{plimpton1995fast}. The system is initially allowed to relax to pressure $P=0$ at constant temperature $T=1.0$; then, $NPT$ simulations are run at these $T,P$ values. An equilibration run of $10^7$ time steps is followed by a production run of $2 \cdot 10^{8}$ time steps. Temperature and pressure are kept constant by Nosé-Hoover chains (three thermostats and barostats) \cite{martyna1994constant,parrinello1981polymorphic,shinoda2004rapid}, insuring a correct sampling of the $NPT$ ensemble. The integration time step is $\delta t=0.003$ for all the simulations, and the relaxation time for the thermostat is chosen to be $\delta t \cdot T_\text{damp} = 0.3$, whereas the relaxation time for the barostat is $\delta t \cdot P_\text{damp} = 3$ \cite{martyna1994constant}. The three dimensions of the box, $L_x,L_y$ and $L_z$, are allowed to fluctuate independently, so that we have access to both volume and shape fluctuations, which are used to estimate the shear and bulk moduli $G$ and $K$, respectively \cite{rovigatti2019connecting}. The total density $\rho$ is thus defined as $\rho\equiv M_\text{tot}/\langle V \rangle$, where $V=L_xL_yL_z$ and $\langle \cdot \rangle$ denotes an average over all configurations. We note that here $M_\text{tot}$ refers to the number of particles after all the dangling ends have been removed (see Tables~\ref{tab:f3} and ~\ref{tab:f4}).

%%%%%%%%%%%%%%%%%%%%%%%%%%%%%%%%%%%%%%%%%%%%%%%%%%%%%%
%%%%%%%%%%%%%%%%%%%%%%%%%%%%%%%%%%%%%%%%%%%%%%%%%%%%%%
\section{Structure}
\label{sec:structure}
%%%%%%%%%%%%%%%%%%%%%%%%%%%%%%%%%%%%%%%%%%%%%%%%%%%%%%
%%%%%%%%%%%%%%%%%%%%%%%%%%%%%%%%%%%%%%%%%%%%%%%%%%%%%%

In this Section we study the topological and structural properties of the networks.

%%%%%%%%%%%%%%%%%%%%%%%%%%%%%%%%%%%%%%%%%%%%%%%%%%%%%%
\subsection{Density and strand length distribution}
\label{subsec:strand_length}
%%%%%%%%%%%%%%%%%%%%%%%%%%%%%%%%%%%%%%%%%%%%%%%%%%%%%%

\begin{figure}
\centering
\includegraphics[width=0.45 \textwidth]{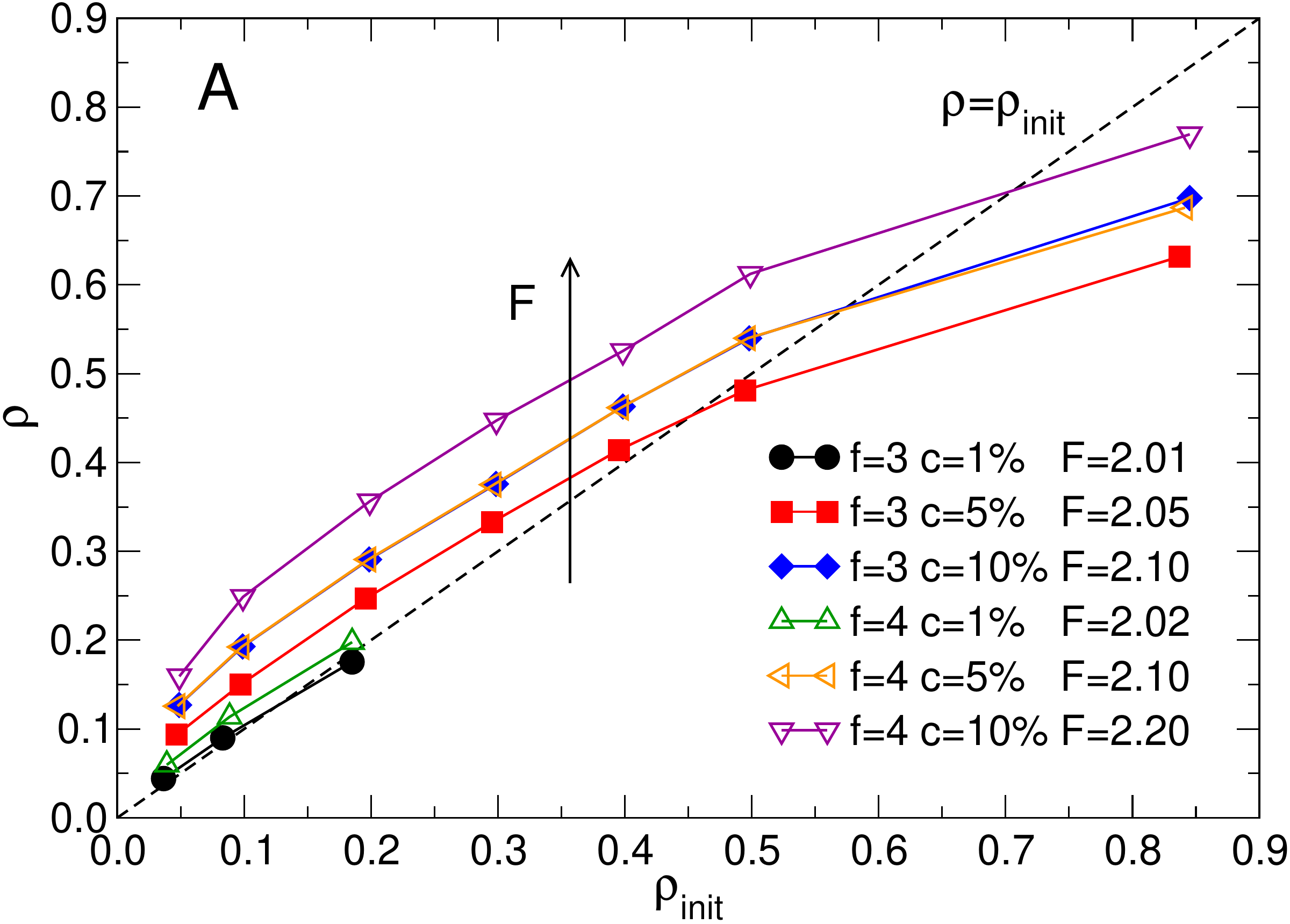}
\includegraphics[width=0.46 \textwidth]{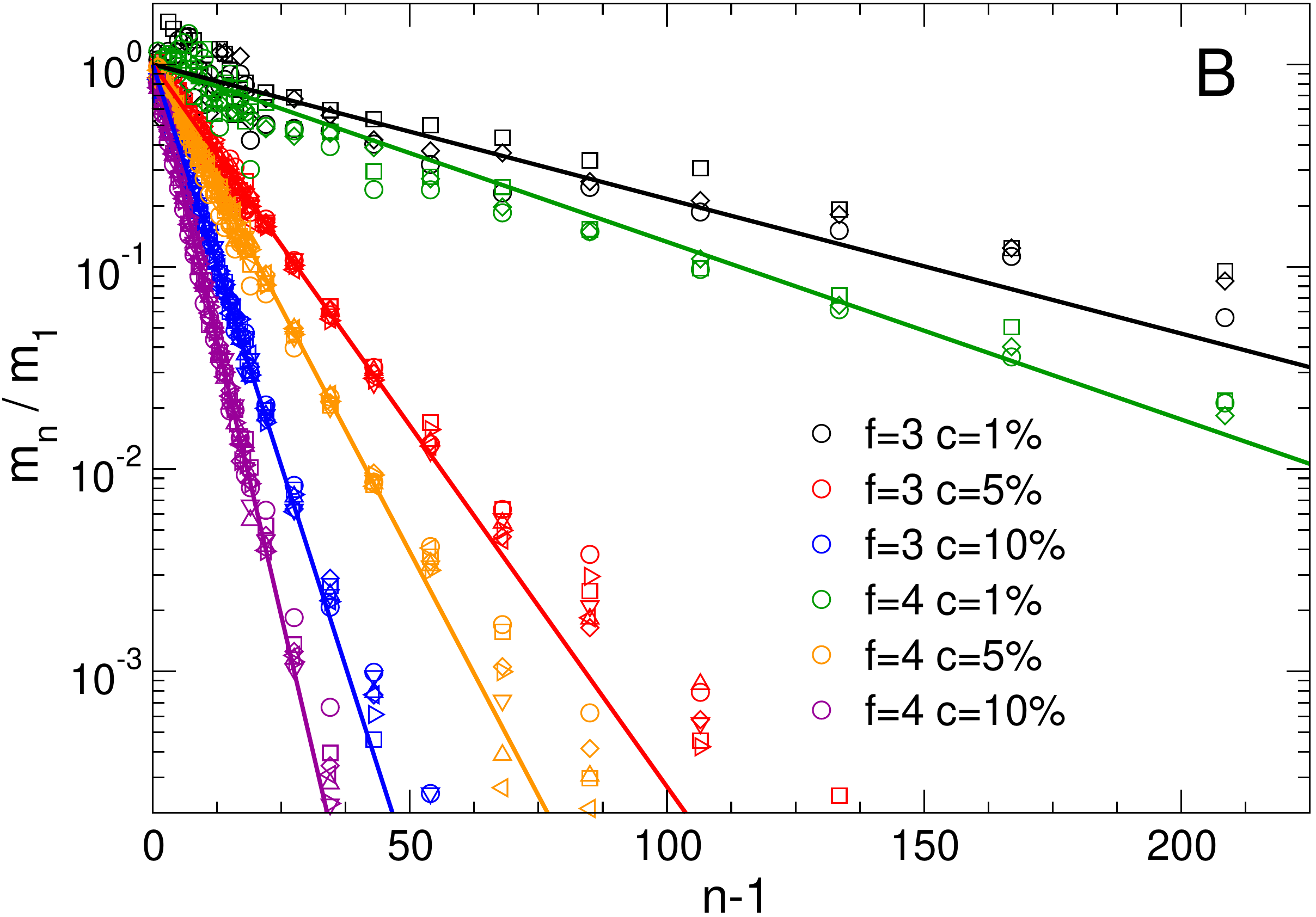}
\caption{(A) Average density, $\rho\equiv M_\text{tot}/\langle V \rangle$, as a function of the assembly density of the system, for different values of $f$ and $c$. $F$ is the mean valence. The $f = 3$, $c = 10\%$ and $f = 4$, $c = 5\%$ curves, which have the same $F$, overlap almost perfectly. (B) Number of strands of length $n$, $m_n$, normalized by the number of strands of unitary length, $m_1$. Points are simulation data, with different colors corresponding to different $f$ and $c$ values and different symbols corresponding to different densities. Lines are the predictions of Eq.~\eqref{eq:length_dist}.}
\label{fig:strand_length}
\end{figure}

After the relaxation at $P = 0$, the final mean density of the system, $\rho \equiv M_\text{tot}/\langle V \rangle$, will in general be different from the assembly density $\rho_\text{init}$. In Fig.~\ref{fig:strand_length}A, we show $\rho$ as a function of $\rho_\text{init}$ for different values of $c$ and $f$. One can see that for most of the systems with $\rho_\text{init}<0.85$, we have $\rho>\rho_\text{init}$, \textit{i.e.}, the network contracts after switching the interaction from patchy to Kremer-Grest and letting the system relax to $P=0$. This is due to the fact that the chains are stiffer with the patchy potential, due to the directionality of the bonds, which is absent with the Kremer-Grest potential. The only exceptions are $f=3,c=5\%,\rho_\text{init}=0.5$ and $f=3,c=1\%,\rho_\text{init}=0.2$, for which the network expands slightly ($\rho < \rho_\text{init}$). For $\rho_\text{init}=0.85$, on the other hand, all the systems expand, as discussed below. Finally, we note that the curves for $f=4,c=5\%$ and $f=3,c=10\%$ superimpose perfectly, implying that the parameter controlling $\rho$ is the mean valence $F$ (Eq.~\eqref{eq:mean_valence}). Indeed, it can be verified that both these systems have $F = 2.1$ (see also Tabs.~\ref{tab:f3} and \ref{tab:f4}). This is expected for systems of patchy particles, where the mean valence $F$ is known to control the equilibrium thermodynamic properties of the system \cite{zaccarelli2005model,bianchi2006phase,sciortino2011reversible,rovigatti2013computing}.

One of the most relevant properties of the network is the strand length distribution, where a strand is a segment of bivalent particles between two crosslinks. Here we define the strand length $n$ as the average number of bivalent particles contained in such segment, so that a strand of length $n$ will comprise of $n$ beads and $n+1$ bonds. We recall that, since two crosslinks cannot bind to each other in our system, the minimum chain length is $n=1$, corresponding to $2$ bonds. Defining $m_n$ as the number of network strands of length $n$, we can compute the normalized distribution of the chemical lengths $n$ of the chains, $m_n/m_1$, which is shown in Fig.~\ref{fig:strand_length}B for systems with different $f,c$ and $\rho$. For all systems, the distribution is exponentially decaying, consistent with the behavior found for random-crosslinking from a melt of precursor chains~\cite{grest1990statistical,higgs1988polydisperse}, and divalent self-assemblying systems~\cite{sciortino2007self,sciortino2008growth,zhang2022dispersity}. Moreover, $m_n$ is idependent of density, as one expects given the equilibrium nature of the assembly protocol. The distribution is given by the well-known formula of Flory \cite{flory1953principles}:

\begin{equation}
\frac{m_n}{m_1} = \left( 1 - \frac{1}{N_s} \right)^{n - 1},
\label{eq:length_dist}
\end{equation}

\noindent
where $N_s$ is the mean strand length, defined as

\begin{equation}
 N_s \equiv \langle n \rangle_n  = \frac 1 {M_s} \sum_{n=1}^\infty m_n n = \frac 2 f \left(c^{-1}-1\right).
 \label{eq:ns}
\end{equation}

\noindent
In Eq.~\eqref{eq:ns}, $\langle \cdot \rangle_n$ denotes over the configurations and over the strands and $M_s$ the total number of strands. The theoretical probability distribution is shown in Fig.~\ref{fig:strand_length}B by continuous lines, and reproduces well the simulation data.

%%%%%%%%%%%%%%%%%%%%%%%%%%%%%%%%%%%%%%%%%%%%%%%%%%%%%%
\subsection{Strand conformation and entanglements}
%%%%%%%%%%%%%%%%%%%%%%%%%%%%%%%%%%%%%%%%%%%%%%%%%%%%%%

\begin{figure}
\centering
\includegraphics[width=0.45 \textwidth]{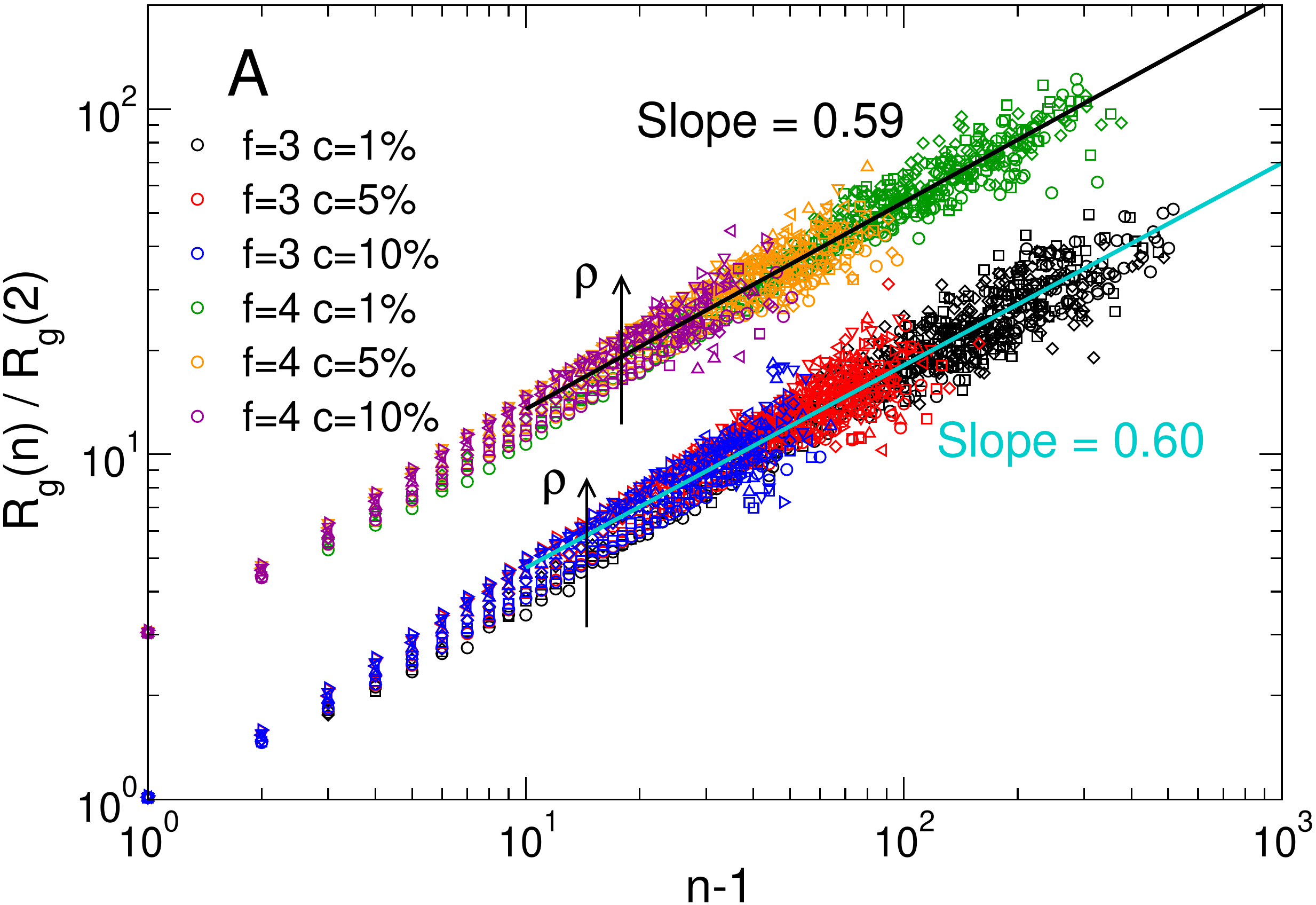}
\includegraphics[width=0.46 \textwidth]{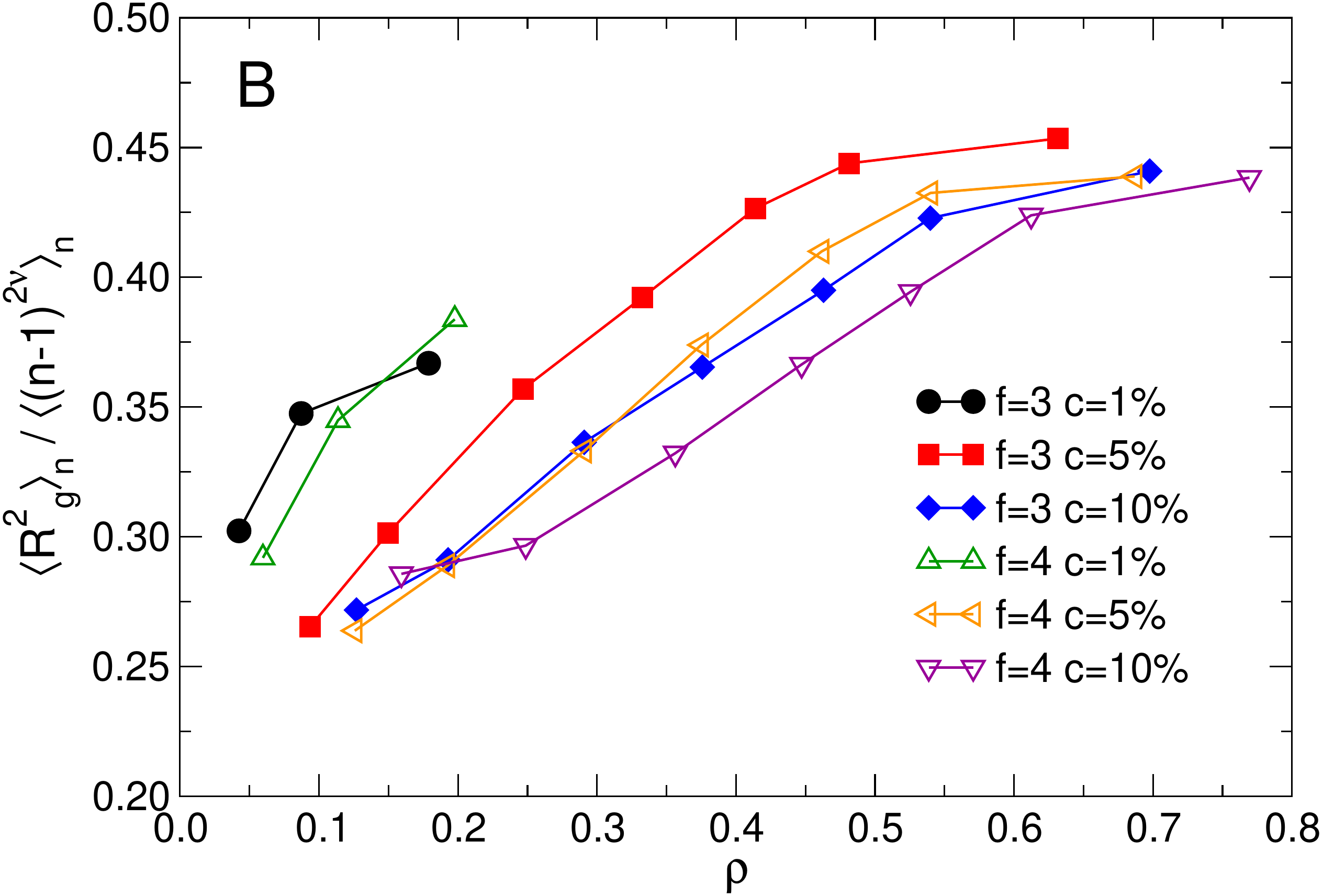}
\caption{(A) Normalized mean radius of gyration of the strands of length $n$ as a function of $n-1$. Points are simulation data, with different colors corresponding to different $f$ and $c$ values and different symbols corresponding to different densities. Solid lines are power-law fits in the range $n-1\geq 10$. The data for $f=4$ have been shifted up by a factor $3$ in order to aid visualization. Note that for each pair of $f,c$ values, all the densities $\rho$ considered are included in the plot, and the vertical arrows denote the direction in which $\rho$ increases. (B) Mean-squared strand radius of gyration $R_g^2\equiv  \langle R_g^2(n) \rangle_{n}$, normalized by $\langle (n-1)^{2\nu} \rangle_n$, as a function of density, for different values of $f$ and $c$.}
\label{fig:rg_50k}
\end{figure}

In order to study the spatial conformation of the strands, we consider the mean radius of gyration of single strands of length $n$, $R_g(n)$. To calculate $R_g$, we consider as part of the strand only the bivalent particles and the bonds that connect them, excluding the crosslinks. Thus, $R_g(1)=0$, since for $n=1$ we only have a single bead, and for $n=2$ (two beads) we have $R_g(2)=l_b/2$, with $l_b$ the bond length. In Fig.~\ref{fig:rg_50k}A we show $R_g(n)/R_g(2)$ for all the simulated systems. As a consequence of Eqs.~\eqref{eq:length_dist} and \eqref{eq:ns}, networks with low $c$ contain on average longer strands. In particular, for $c=1\%$ we observe chain lengths $n$ up to $\simeq 500$. Fitting $R_g(n)$ for all $c$ values to the power law $(n-1)^{\nu}$ in the range $n-1\geq 10$, yields $\nu=0.59 \pm 0.02$ for $f=3$ and $\nu=0.60 \pm 0.02$ for $f=4$ (see Fig.~\ref{fig:rg_50k}A). Within the accuracy of the data, these values are compatible with the Flory exponent for self-avoiding walks, \textit{i.e.}, $\nu=0.59$ \cite{rubinstein2003polymer}, suggesting that in the considered range of $n$, the strands adopt conformations akin to those of linear chains in a good solvent. The inverse of the above metric exponent yields the fractal dimension of the individual strands,  $1/\nu=1.7$. Note that this fractal dimension is, in general, different from that of the whole network, and that for randomly branched polymers this is predicted to be $d_f=2$ \cite{parisi1979hausdorff}. We also note that at fixed $n$, $R_g(n)$ is basically independent of the crosslink concentration $c$ and of the crosslink valence $f$, and depends only weakly on the density of the system. This is shown in Fig.~\ref{fig:rg_50k}B, where we plot the quantity $\langle R_g^2(n) \rangle_{n}/\langle (n-1)^{2\nu} \rangle_n$, which is proportional to the effective Kuhn length \cite{rubinstein2003polymer} of the chains. For all systems considered here, this is a monotonically increasing function of $\rho$ and, discounting the numerical noise, a monotonically decreasing function of $c$. From Fig.~\ref{fig:rg_50k}B, we also see that in the systems $f=4,c=5\%$ and $f=3,c=10\%$, which have the same mean valence $F$, the strands have the same conformation. The same qualitative behavior is observed when considering the end-to-end distance $R_e$ (not shown).

\begin{figure}
\centering
\includegraphics[width=0.45 \textwidth]{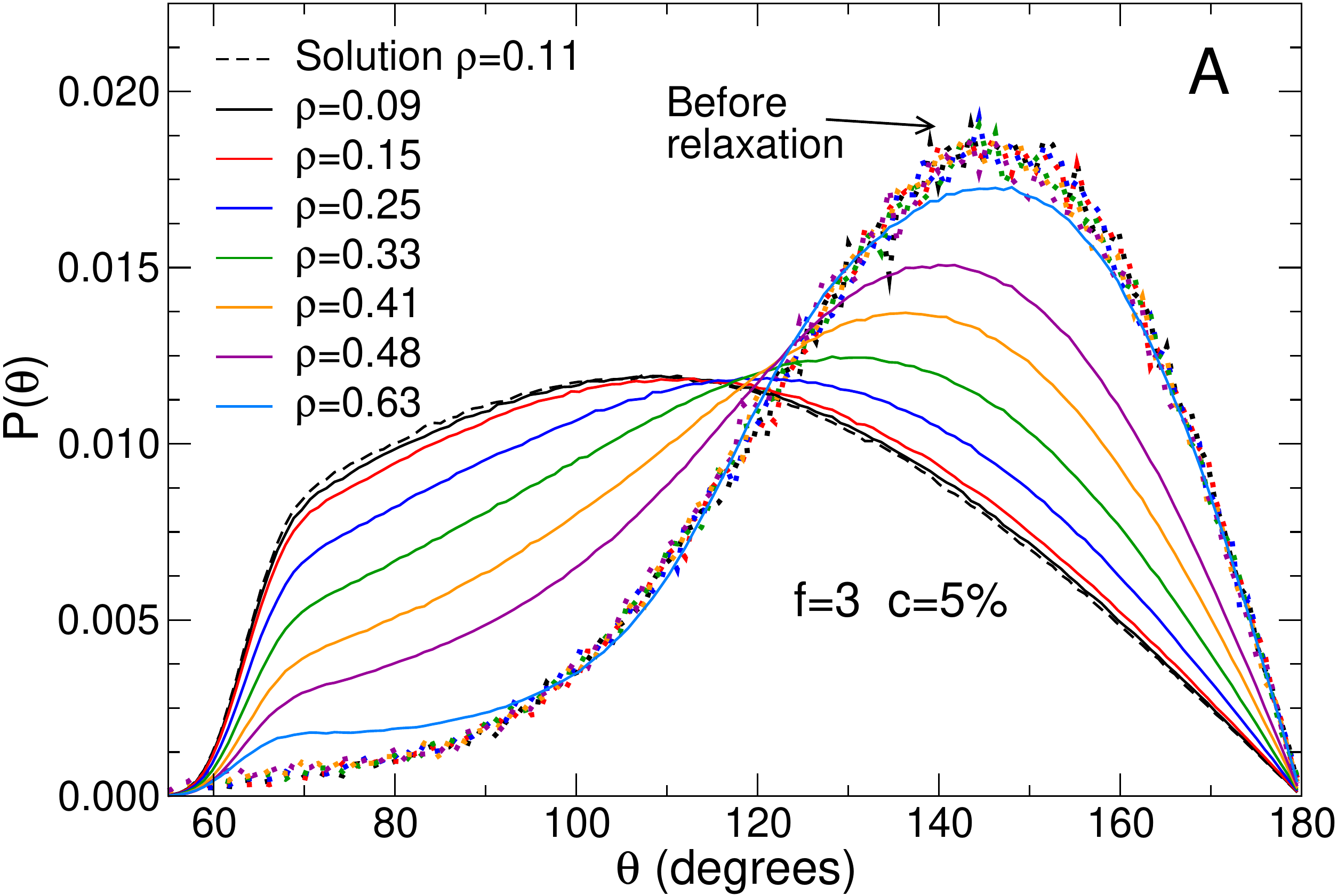}
\includegraphics[width=0.45 \textwidth]{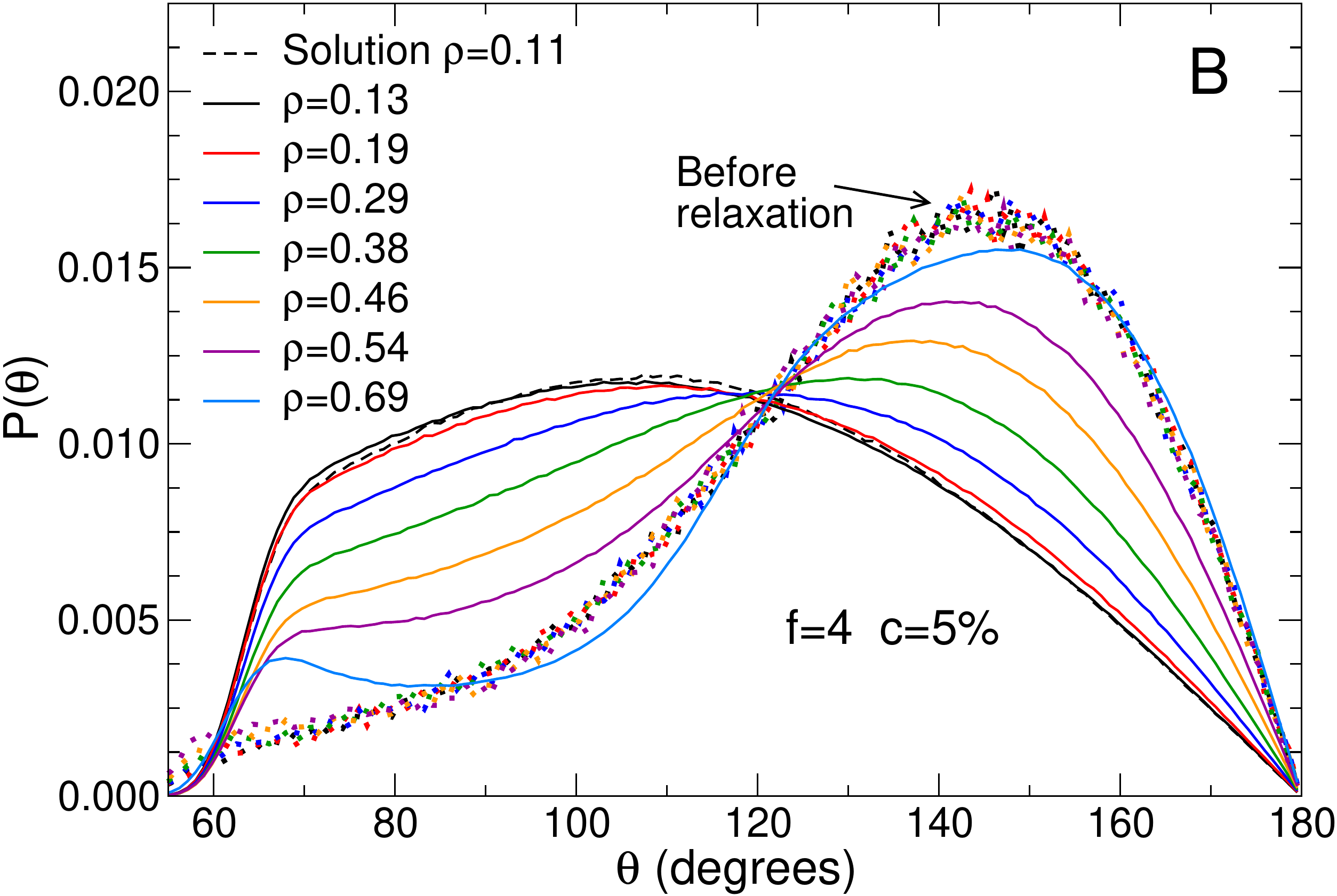}
\caption{Bond angle distribution $P(\theta)$ for $f=3,c=5\%$ (A) and $f=4,c=5\%$ (B). The dotted lines are bond angle distributions before the system is allowed to relax to zero pressure. Dashed lines: data for a polymer solution of $50$ chains of length $1000$ for $\rho=0.11$ (data from Ref. \cite{sorichetti2020determining}).}
\label{fig:bond_angle_c5}
\end{figure}

The fact that $R_g$ increases with density might surprise, since in polymeric systems with free chains, $R_g$ usually \emph{decreases} with increasing $\rho$ \cite{rubinstein2003polymer}. This behavior, however, can be understood qualitatively by analyzing the bond angle distribution of the strands, $P(\theta)$, where $\theta$ is the angle between bonded triplets of monomers. In Fig.~\ref{fig:bond_angle_c5} we report with solid lines $P(\theta)$ for $c=5\%$ and $f=3$ (A) and $f=4$ (B) (the other systems show the same qualitative behavior). These are compared with the distribution obtained before changing the interaction potential from patchy to Kremer-Grest and allowing the system to relax to $P=0$ (dotted lines). Before the system is brought to $P=0$, $P(\theta)$ displays a peak at $\theta \simeq 142^\circ$, which originates from the the patchy potential. Moreover, $P(\theta)$ drops to zero for $\theta\lesssim 60^\circ$, which is the minimum allowed angle considering the excluded volume (WCA) interaction.

We note that before relaxation $P(\theta)$ only depends on $f$ and $c$ and not on $\rho_\text{init}$. As the bonding potential is switched to the directional patchy one to the non-directional FENE one and the system relaxes to $P=0$, the curve becomes broader, and the peak shifts to lower values of $\theta$. At low $\rho$ the system can completely relax and $P(\theta)$ assumes a form identical to that of chains in a dilute solution (dashed lines; data from Ref.~\cite{sorichetti2020determining}). By contrast, for high $\rho$, $P(\theta)$ retains a shape which is quite similar to the shape it had before relaxation, signaling the presence of strong topological constraints, \textit{i.e.}, entanglements \cite{rubinstein2003polymer}, resulting from the non-crossability of the strands. These constraints prevent the bond angle from fully relaxing, as signaled by the survival of the $\theta \simeq 142^\circ$ peak. We can see that for both systems, the average bond angle $\langle \theta \rangle$ increases with increasing density: This results in an increase of the effective persistence length \cite{rubinstein2003polymer} of the strands, and therefore an increase of $R_g(n)$, as shown in Fig.~\ref{fig:rg_50k}B. This phenomenon is reminiscent of topological rigidification, for which the system's rigidity increases with increasing topological complexity as observed in knotted rings and star polymers~\cite{vargas2018communication}, although the limited range of strand lengths available here does not allow to confirm the exact nature of this rigidification. We note that the persistence length of chains of patchy particles in solution decreases with $\rho_\text{init}$, but only weakly (see Appendix~\ref{app:details}); thus, this  $\rho_\text{init}$-dependence cannot, by itself, explain the effect observed in $P(\theta)$.

\begin{figure}
\centering
\includegraphics[width=0.45\textwidth]{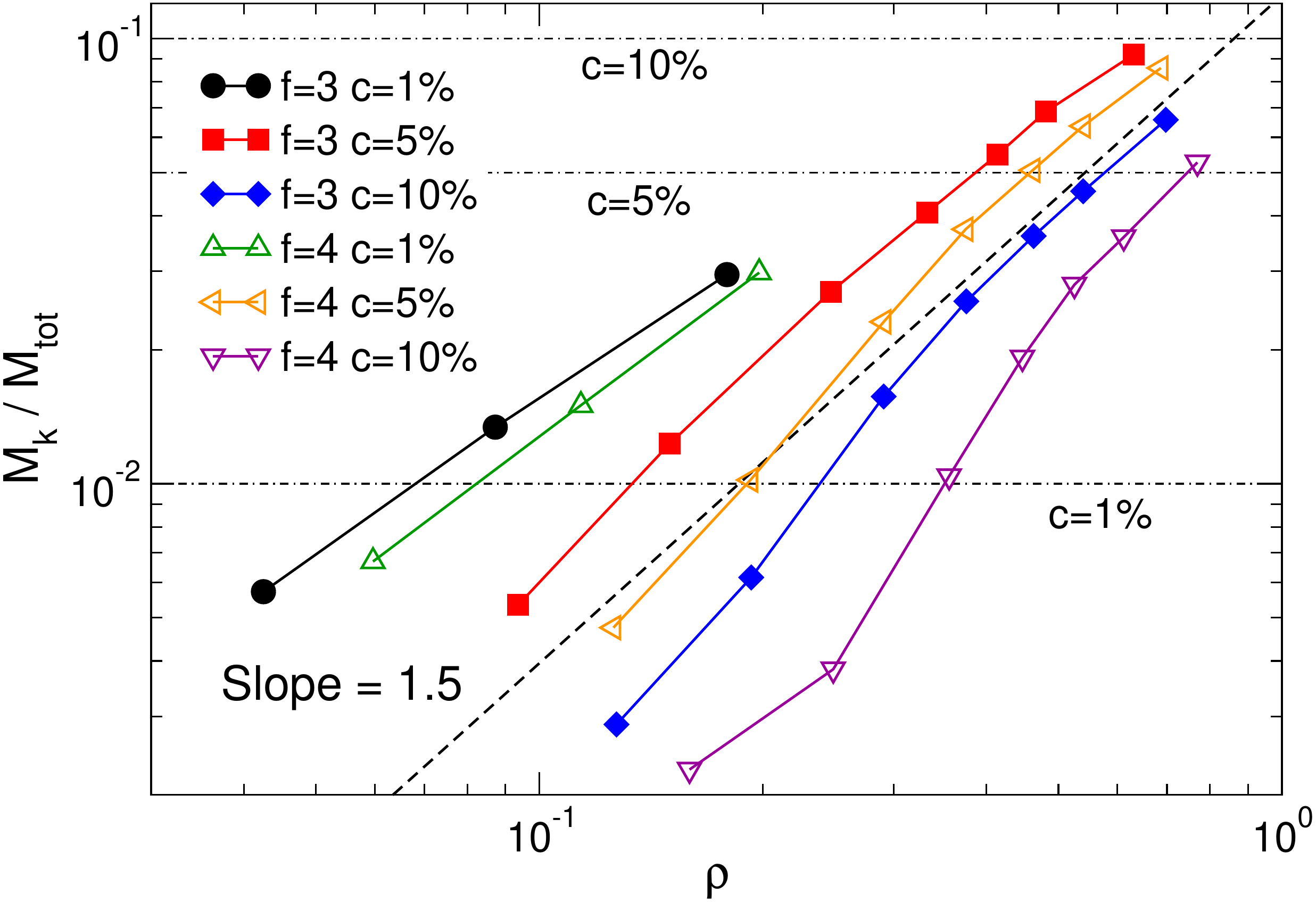}
\caption{Number of kinks $M_k$ (obtained from primitive path analysis), normalized by the total number of monomers $M_\text{tot}$, as a function of monomer density $\rho$, for different values of $f$ and $c$. Dashed line: power law with slope $1.5$. Dash-dotted lines: the three crosslink fractions investigated ($c=1\%,5\%,$ and $10\%$).}
\label{fig:entanglements}
\end{figure}

The increase of the number of entanglements with increasing density can be more directly quantified using the method of \textit{primitive path analysis} \cite{everaers2004rheology,sukumaran2005identifying}. The procedure is as follows: (1) The crosslinks are fixed in space, then (2) the intra-chain excluded volume interactions are turned off, while the inter-chain interactions are kept, and finally (3) the system is cooled to $T=0$. Since intra-chain excluded volume interactions are turned off, the chains are straightened; however, the inter-chain interaction still prevents the chains from passing through each other, and thus the topology is conserved. The network is thus reduced to a mesh of primitive paths. We then count the number of contacts between any two primitive paths, and refer to it as the the number of kinks, $M_k$. Under the assumption that two strands do not entangle more than once, $M_k$ is proportional to the number of entanglements $M_e$, \textit{i.e.}, $M_k \propto M_e \simeq M_\text{tot} / 4 N_e$, where $N_e$ is the entanglement length \cite{rubinstein2003polymer,everaers2004rheology}. One can see from Fig.~\ref{fig:entanglements} that the fraction of kinks $M_k / M_{\rm tot}$ increases with $\rho$ and $f$ and decreases with $c$. The $\rho$-dependence at high-density seems to be compatible with a power-law of exponent $1.5$, although the limited density range does not allow to draw any definite conclusion (see the dashed line reported in Fig.~\ref{fig:entanglements} as reference). It is important to stress that in our systems an increase in $\rho$ is not equivalent to that which one would obtain by simply compressing the system (\textit{i.e.}, by collapsing the network), since the degree of entanglement of the network strongly depends on the assembly density $\rho_\text{init}$.

The results of the primitive path analysis presented above is important for several reasons: First of all, from the methodology point of view it is clear that in the protocol we employ to build disordered networks the number of topological kinks (and therefore of entanglements) can be controlled, in a statistical sense, by varying $\rho_{\rm init}$. Leveraging this feature will make it possible to investigate fully-entangled, polydisperse, disordered networks. Secondly, for the particular choice of parameters we made here, these results show that the investigated systems are at most lightly entangled, as the order of magnitude of the number of entanglements (which is smaller than $M_k$) never exceeds that of the number of crosslinks (also shown in Fig~5). Finally, the increase of the number of entanglements with density may explain the apparent topological rigidification discussed above.

%%%%%%%%%%%%%%%%%%%%%%%%%%%%%%%%%%%%%%%%%%%%%%%%%%%%%%
\subsection{Structure factor}
\label{subsec:network_sq}
%%%%%%%%%%%%%%%%%%%%%%%%%%%%%%%%%%%%%%%%%%%%%%%%%%%%%%

\begin{figure}
\centering
\includegraphics[width=0.45 \textwidth]{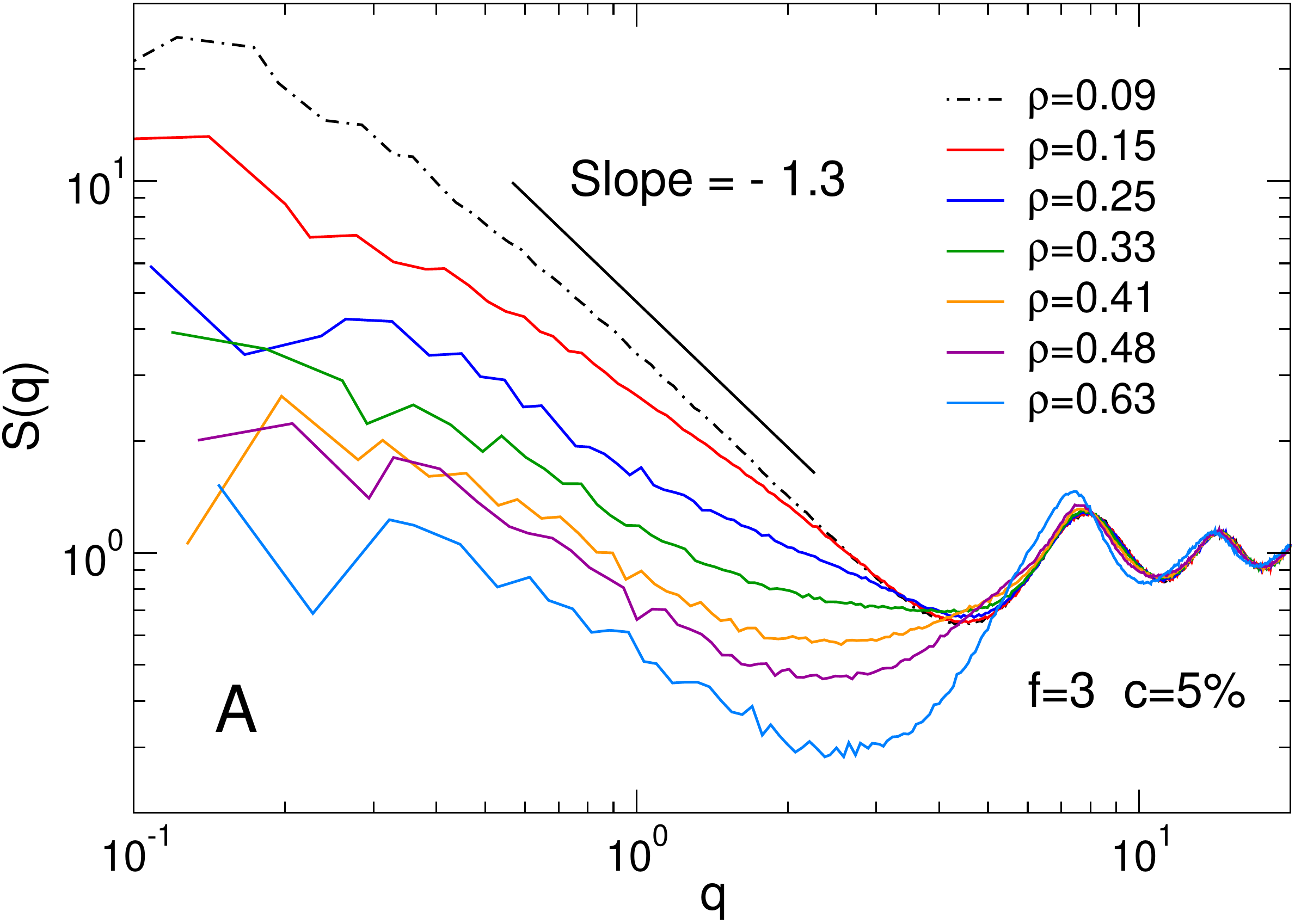}
\includegraphics[width=0.45 \textwidth]{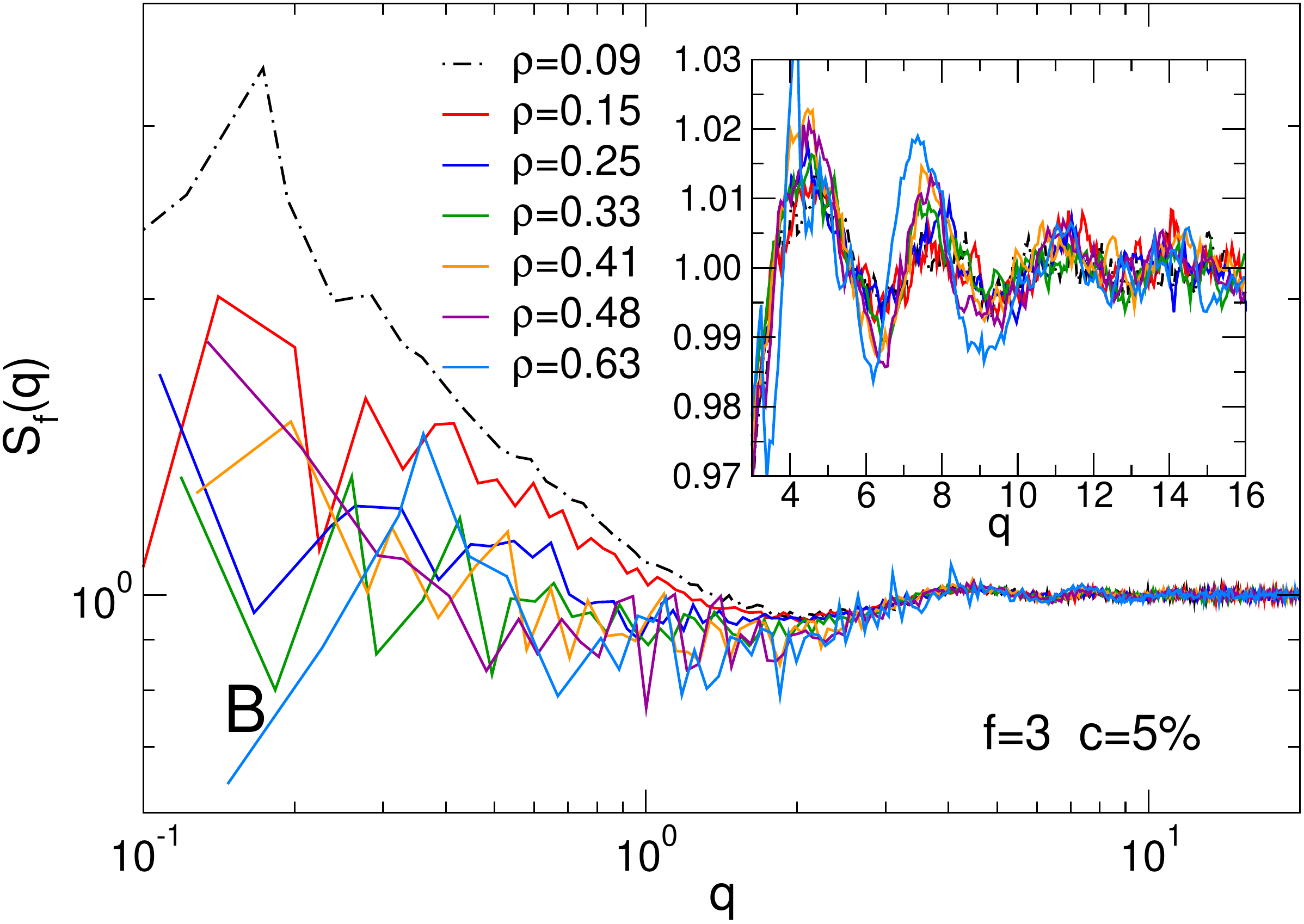}
\includegraphics[width=0.45 \textwidth]{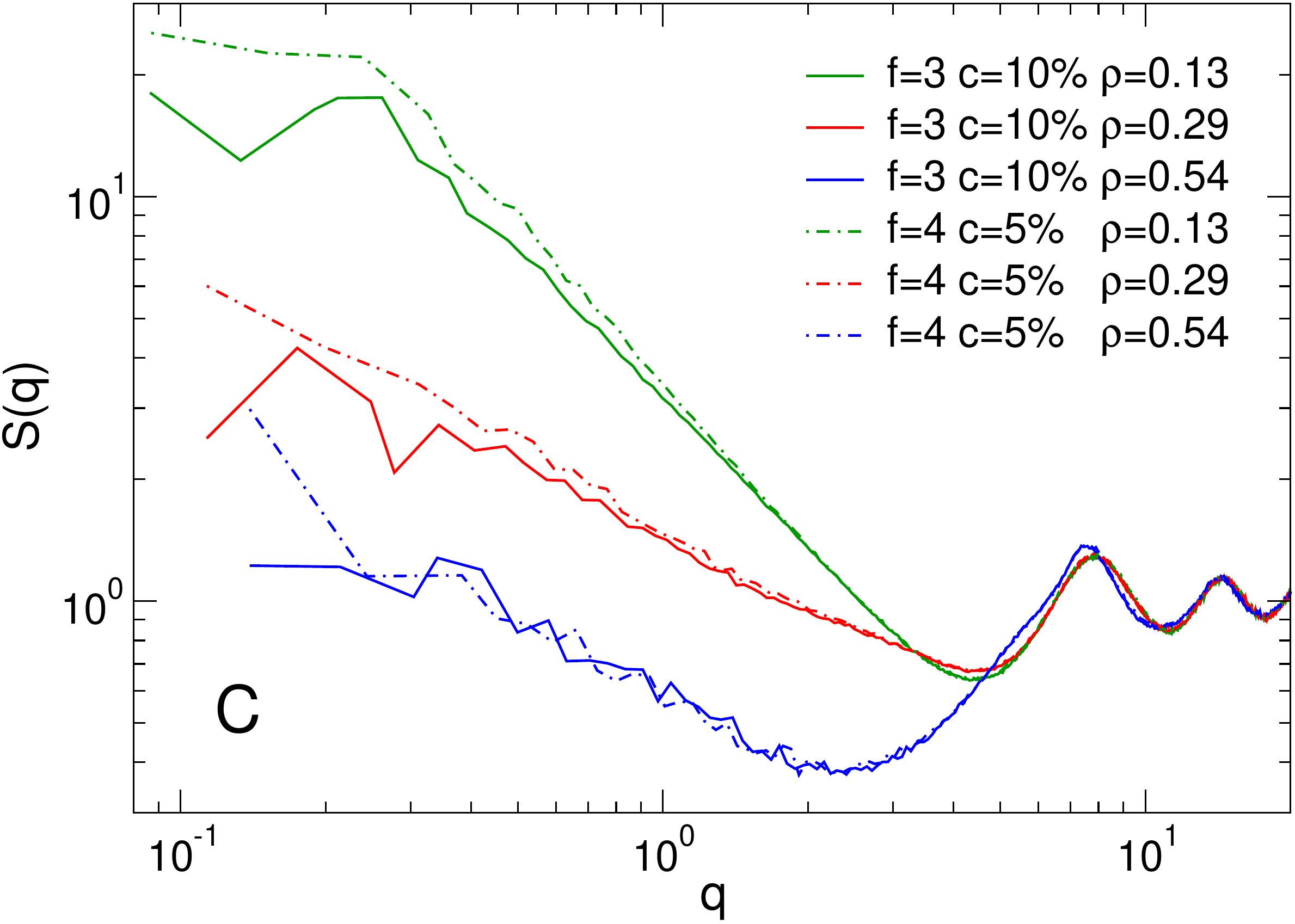}
\includegraphics[width=0.45 \textwidth]{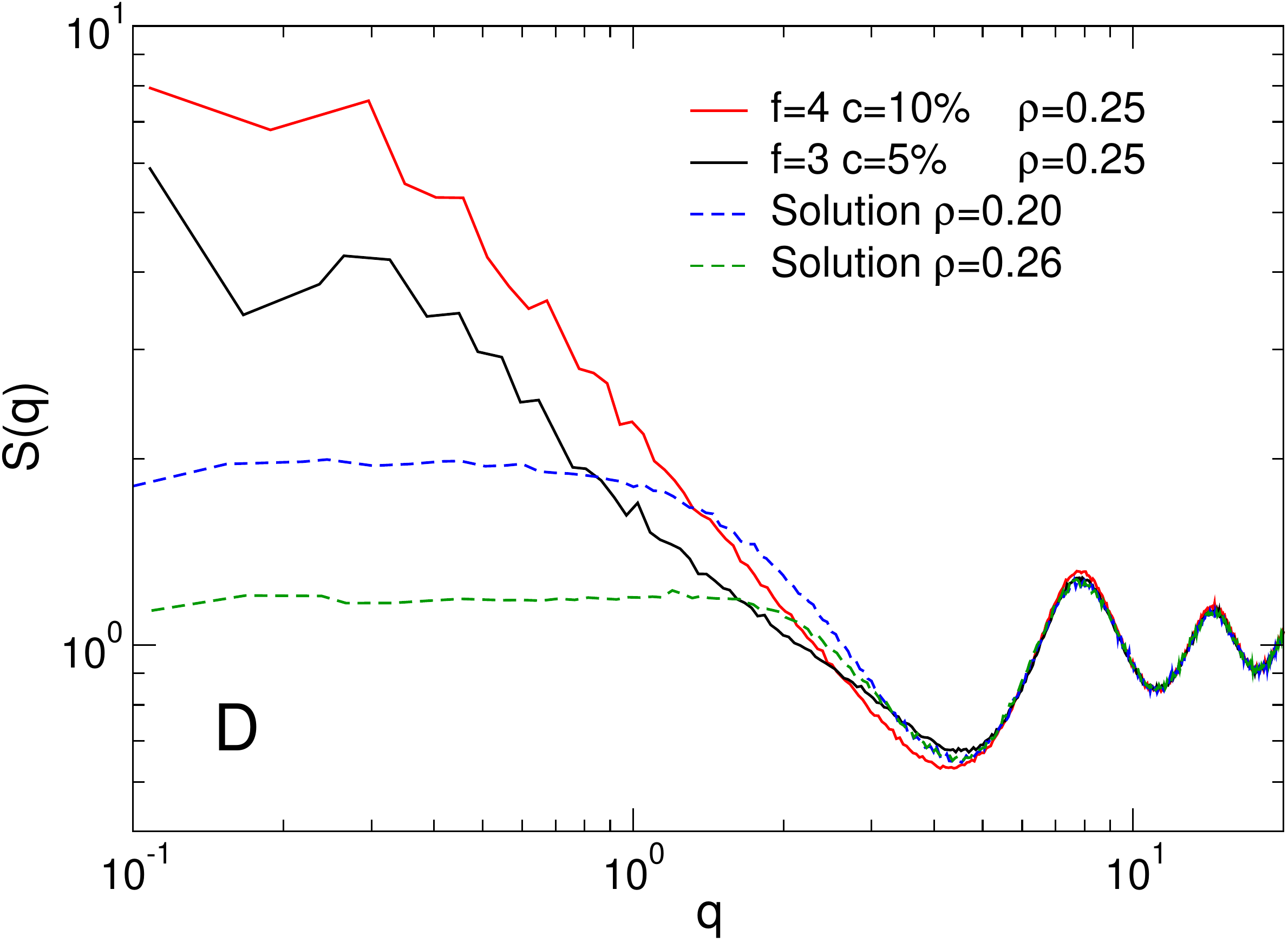}
\caption{(A) Structure factor of all the particles for the system with $f=3,c=5\%$. At low density, the behavior of $S(q)$ is consistent with a power law with exponent $-\alpha$, consistent with the fractal and polydisperse nature of the network (slope with $\alpha=1.3$ reported for comparison). (B) Structure factor of the crosslinks (same systems as in A).  \textit{Inset}: $S_f(q)$ on linear scale. (C) Comparison between the structure factor $S(q)$ of the $f=3,c=10\%$ system and that of the  $f=4,c=5\%$ system for different densities. Note that these systems have the same mean valence $F$ and the same density $\rho$. (D) Comparison between the structure factor of two selected systems which have the same density $\rho=0.25$, but different $F$ (continuous lines). Dashed lines: data for a polymer solution of $50$ chains of length $1000$ for $\rho=0.20$ and $0.26$ (data from Ref. \cite{sorichetti2020determining}).}
\label{fig:sq}
\end{figure}

To gain insight into the global structure of the network, it is useful to study the structure factor $S(q)$ \cite{hansen1990theory}. We report the total structure factor $S(q)$ in Fig.~\ref{fig:sq}A and the structure factor of the crosslinks $S_f(q)$ in Fig.~\ref{fig:sq}B for $f=3,c=5\%$ and different values of the monomer density $\rho$. The structure factors of the other systems display the same qualitative features. All the systems have rather large isothermal compressibility $\kappa_T=S(0)/\rho k_B T$ \cite{hansen1990theory}, which increases with decreasing density. In a finite range of small wave-vectors $q$, the behavior of $S(q)$ is compatible with a power law, $S(q)\propto q^{-\alpha}$. This behavior results from the fractal structure and inherent (strand-)polidispersity of the network, and also from the presence of holes with a wide size range, clearly visible in Fig.\ref{fig:snapshots}~\cite{roldan2017connectivity,horkay2021comparative,chremos2022molecular}. The value of $\alpha$ decreases with increasing monomer density; for the systems studied here, it was found that $0.5 \lesssim \alpha \lesssim 1.3$.  For larger $q$, $S(q)$ behaves similarly to that of a liquid, with a contact peak at $q\simeq 8 \simeq 2 \pi /\sigma$, followed by periodic oscillations. 

In Fig.~\ref{fig:sq}B, we show the structure factor of the crosslinks, $S_f(q)$, for the same systems. The qualitative behavior of $S_f(q)$ is very similar to that of $S(q)$, with power-law decrease at small $q$, followed at larger $q$ by a liquid-like oscillatory behavior (inset). However, the height of $S_f(q)$ in the small-$q$ limit is much lower than that of $S(q)$, and also the oscillations have much smaller amplitude, showing that the location of the crosslinks is more random than the one of the monomers. Note that since two $f$-valent particles cannot bind, the distance of closest approach between two crosslinks is $\simeq 2\sigma$ and therefore the contact peak of $S_f(q)$ is found at $q \simeq \pi / \sigma$.

In Fig.~\ref{fig:sq}C we compare the total structure factors $S(q)$ of the systems $f=4,c=5\%$ and $f=3,c=10\%$ for different values of the total density $\rho$. As noted above, these two systems have approximately the same mean valence $F$ and therefore also same ratio $\rho/\rho_\text{init}$ (Fig.~\ref{fig:strand_length}A). One can see that curves corresponding to similar densities superimpose almost perfectly, suggesting that $S(q)$ is controlled either by $\rho$ alone or by $\rho$ and $F$. In Fig.~\ref{fig:sq}D we compare $S(q)$ for two systems with very similar densities but rather different $F$ ($2.05$ and $2.20$): The two curves are quite different, suggesting that the behavior of $S(q)$ is determined by both $F$ and $\rho$. These two structure factors are also compared with the $S(q)$ of a solution of $50$ chains of length $1000$ at similar densities (dashed lines; data from Ref. \cite{sorichetti2020determining}), evidencing a rather different low-$q$ behavior. The power-law behavior of the network's $S(q)$ at small $q$ suggests that the structure of the network is significantly more heterogeneous than that of the solution at sufficiently large length scales or, equivalently, the inhomogeneities caused by the crosslinks make the networks much more compressible than their chain-only counterparts.

%%%%%%%%%%%%%%%%%%%%%%%%%%%%%%%%%%%%%%%%%%%%%%%%%%%%%%
%%%%%%%%%%%%%%%%%%%%%%%%%%%%%%%%%%%%%%%%%%%%%%%%%%%%%%
\section{Elasticity}
\label{sec:elasticity}
%%%%%%%%%%%%%%%%%%%%%%%%%%%%%%%%%%%%%%%%%%%%%%%%%%%%%%
%%%%%%%%%%%%%%%%%%%%%%%%%%%%%%%%%%%%%%%%%%%%%%%%%%%%%%

In this Section we investigate the long-time limit of the mean-squared displacement of the particles and connect it to the shear modulus of the networks.

%%%%%%%%%%%%%%%%%%%%%%%%%%%%%%%%%%%%%%%%%%%%%%%%%%%%%%
\subsection{Localization length}
\label{sec:lambda}
%%%%%%%%%%%%%%%%%%%%%%%%%%%%%%%%%%%%%%%%%%%%%%%%%%%%%%

\begin{figure}
\centering
\includegraphics[width=0.45 \textwidth]{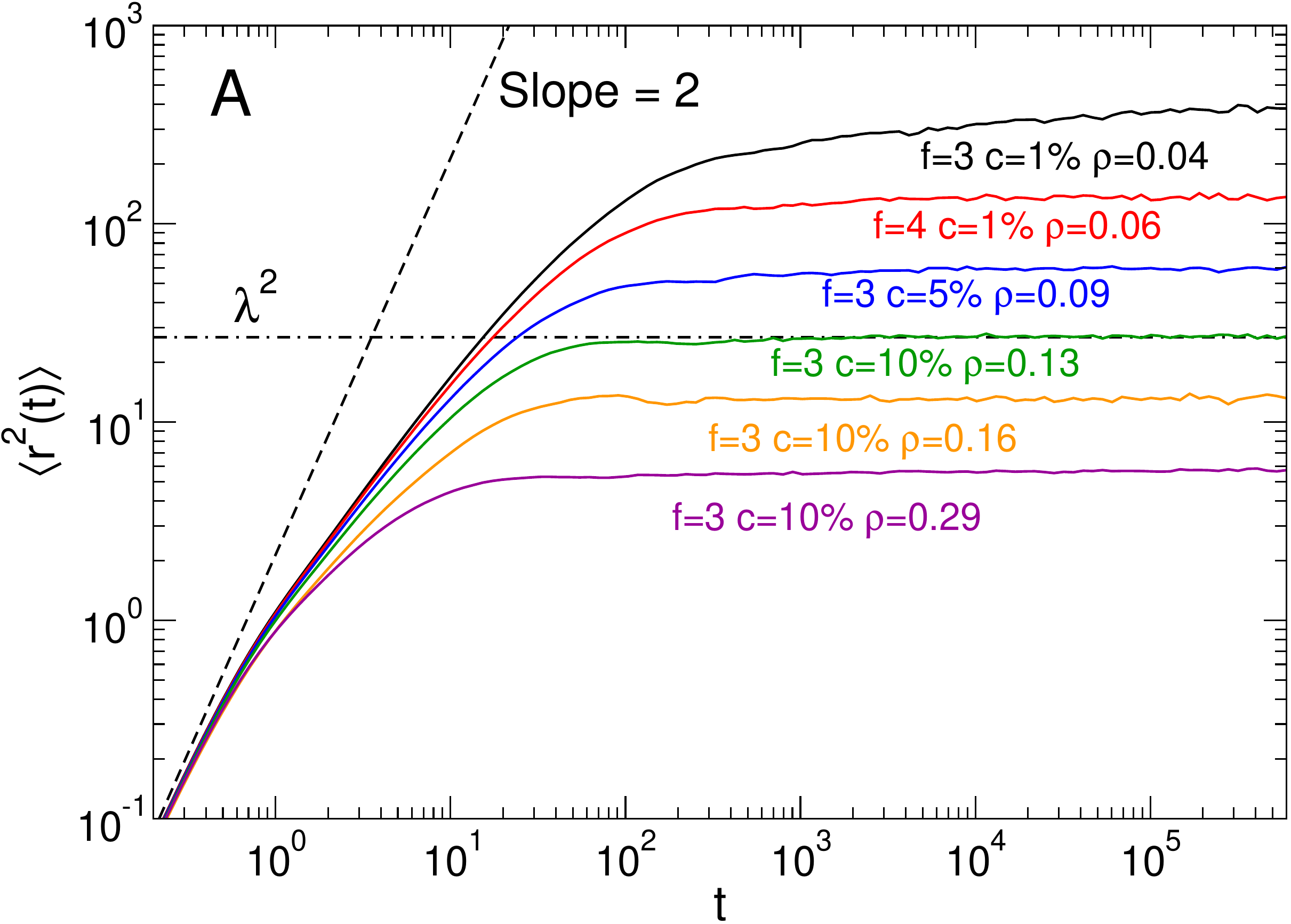}
\includegraphics[width=0.448 \textwidth]{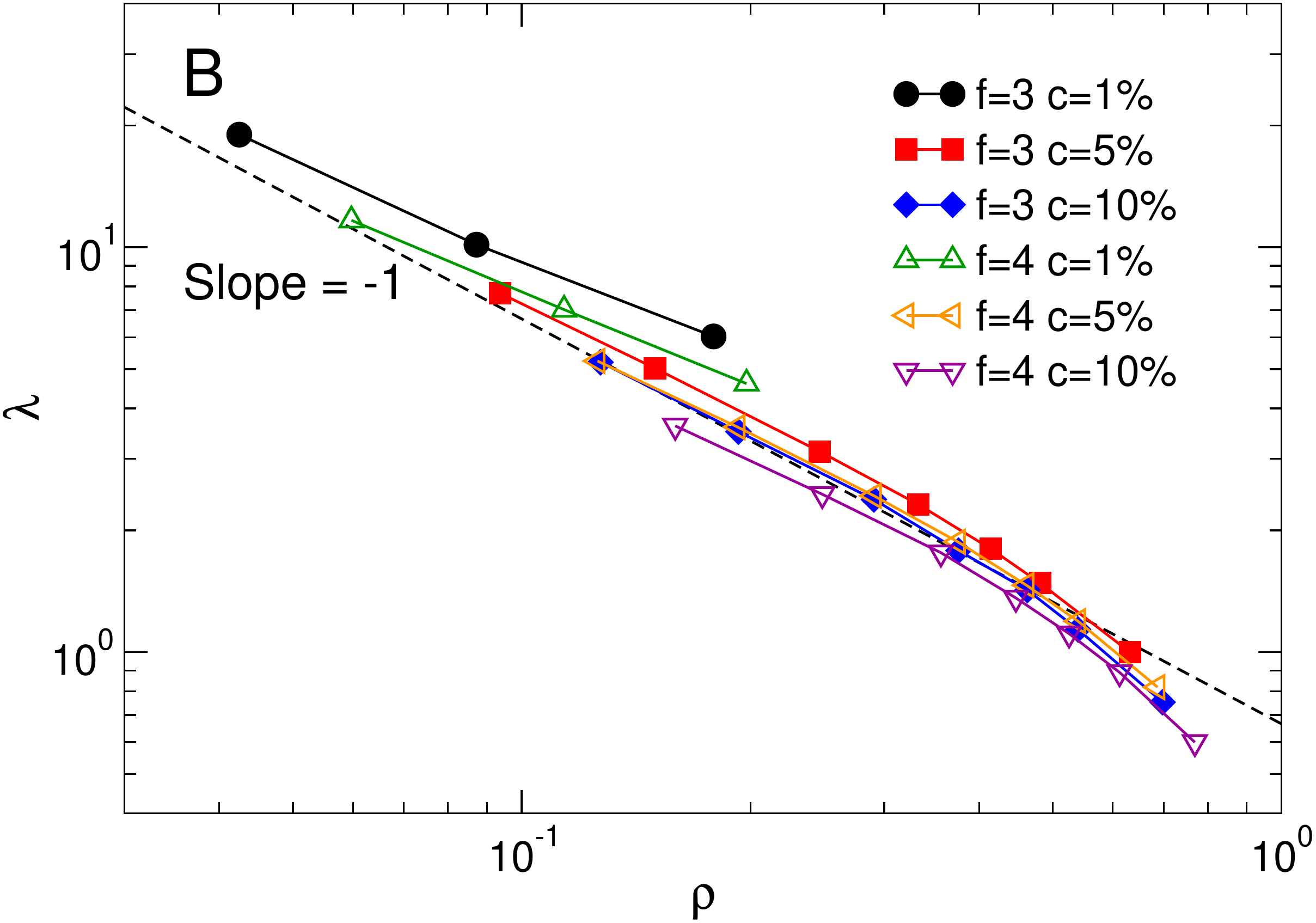}
\caption{(A) MSD of all the particles for some selected systems. Dash-dotted line: the localization length $\lambda^2$, defined as the long-time limit of the MSD (Eq.~\eqref{eq:lambda_def}), for $f=3$, $c=10\%$, $\rho=0.13$ (green line). Dashed line: power law with slope $2$ (ballistic regime). (B) Squared localization length of all the particles as a function of $\rho$ for all investigated systems. Dashed line: power law with slope $-1$.}
\label{fig:msd_lambda}
\end{figure}

To probe the dynamics of the system, we consider the mean-squared displacement (MSD) of the particles, defined as \cite{hansen1990theory}

\begin{equation}
\langle r^2(t) \rangle \equiv {\langle |\mathbf r(t) - \mathbf r(0)|^2 \rangle},
\end{equation}

\noindent 
where $\mathbf r(t)$ is the particle's position vector. Since we run simulations, the simulation box experiences an affine motion due to the change of length of its edges. When computing the MSD we remove this motion (which is almost negligible due to the large system size). In Fig.~\ref{fig:msd_lambda}A we show the MSD of all the particles for some selected systems, which are representative of the general behavior. After the initial ballistic regime, $\langle r^2(t) \rangle \propto t^2$ (dashed line), the MSD crosses over towards a plateau, a behavior which is typical of crosslinked networks and characterizes the system as an elastic solid \cite{duering1994structure,zaccarelli2005model,delgado2005structure,russo2009reversible}. At high density, this crossover is almost immediate, whereas at low density an intermediate superdiffusive regime, $\langle r^2(t) \rangle \propto t^\beta$, with $1<\beta<2$ is observed, resulting from the free motion of chains with different lengths. The square root of the MSD in the plateau region, 

\begin{equation}
\lambda \equiv \left[ \lim_{t \to \infty} \langle r^2 (t) \rangle \right]^{1/2},
\label{eq:lambda_def}
\end{equation}

\noindent
is a \textit{static property} of the system that is known as the \emph{localization length}, which is an important quantity in many systems~\cite{kumar2001gelation,zaccarelli2005model,demichele2011scaling} and corresponds to the typical length scale of the motion the particles perform around their equilibrium positions. One sees that $\lambda$ increases with decreasing $\rho$, going from $\lambda \simeq 1$ for $\rho=0.29$ to $\lambda \simeq 20$ for $\rho=0.04$. These values can be compared with those typical of glass-forming liquids, where $\lambda \simeq 0.1$ due to the localization being mainly determined by local packing constraints \cite{kob1995testing,binder2011glassy}. By contrast, here the localization stems from topological constraints, namely the connectivity of the network and the non-crossability of the chains, and the observed plateau is not transient, but linked to the rubbery modulus. Therefore, the localization length we study is different from the one typical of glassy segmental dynamics~\cite{zheng2022understanding}. The topology of the system is fixed, but the particles can still participate in large amplitude oscillations involving substantial parts of the network \cite{zaccarelli2005model,delgado2005structure,rovigatti2011vibrational}. We note that at low densities the approach to the plateau occurs at very long time scales, since even in the absence of dangling ends these networks can have extremely long relaxation times, due to the long time it takes the longest strands to explore the whole configuration space. Despite this, it is possible to estimate $\lambda$ with good accuracy for all the studied systems. 

In Fig.~\ref{fig:msd_lambda}B we show the localization length as a function of the monomer density $\rho$. In the density we studied, the dependence of $\lambda$ on $\rho$ is roughly compatible with a power law with exponent $-1$, \textit{i.e.}, $\lambda \propto \rho^{-1}$ although the curves display a steeper decrease at the highest densities, where local packing starts to play a more significant role. The same behavior is observed when considering the localization length of the crosslinks or that of the bivalent particles, see Fig.~\ref{fig:lambda_all} in the Appendix. We note that also in this case the curves for $f=4, c=5\%$ and $f=3, c=10\%$, which have the same mean valence $F$, superimpose almost perfectly, confirming once more that the static properties of the system are controlled by $F$ alone, as also shown in Sec.~\ref{sec:structure}. As the strand length distribution is independent of density, we ascribe the decrease of $\lambda$ with $\rho$ to topological constraints stemming from the non-crossability of the chains, and possibly to a difference in the overall network topology.

\begin{figure}
\centering
\includegraphics[width=0.45 \textwidth]{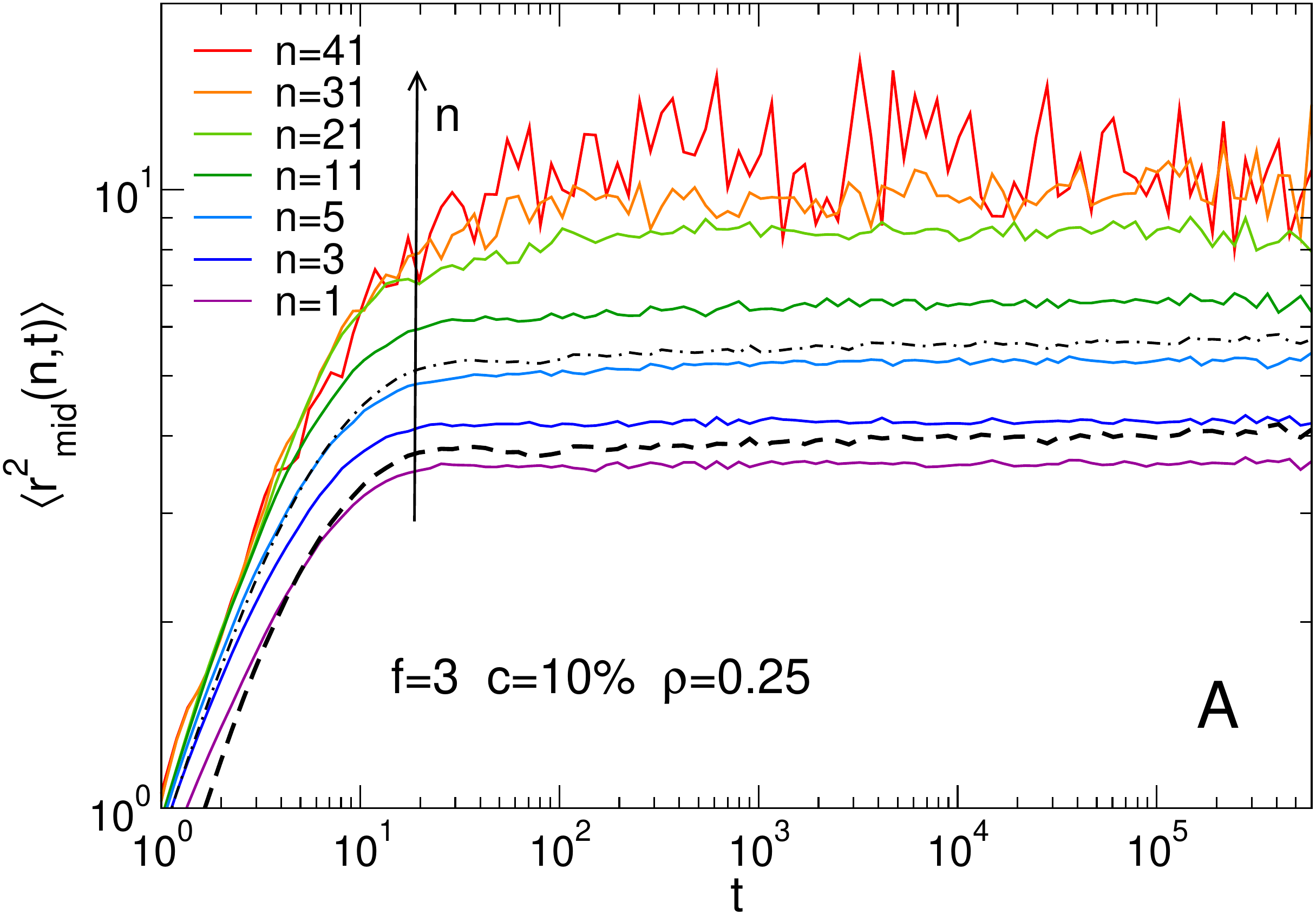}
\includegraphics[width=0.45 \textwidth]{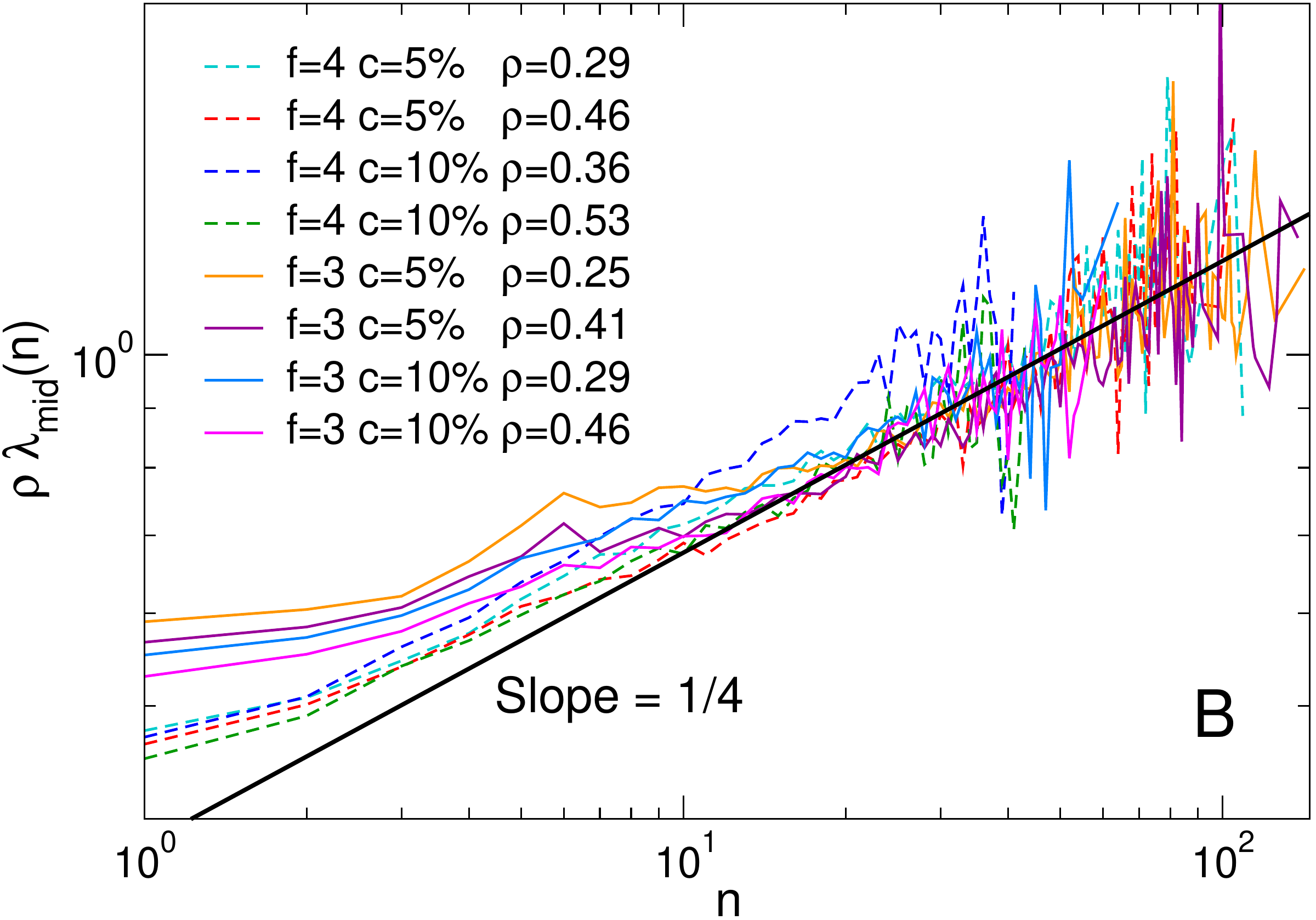}
\caption{(A) MSD of the middle monomers of the strands of length $n$ for $f=3,c=10\%,\rho=0.25$. Dashed lines: MSD of the crosslinks. Dash-dotted lines: MSD of all the particles. (B) localization length of the central monomers of the strands $\lambda_\text{mid}(n)$, multiplied by the monomer density $\rho$, as a function of the strand length.}
\label{fig:middle_plateau_400k}
\end{figure}

To better understand how the entanglements affect strands of different lengths, we consider the MSD of the central monomers of odd-length strands, $\langle r_\text{mid}^2 (n,t) \rangle$, shown in Fig.~\ref{fig:middle_plateau_400k}A for $f=3, c=10\%, \rho=0.25$. In order to improve the statistics, we show here the data for systems of $4\cdot 10^5$ particles instead of those of $5\cdot10^4$. We observe that the localization length $\lambda_\text{mid}(n)$ of the central monomers, which for $n=1$ is very similar to that of the crosslinks (dashed line), increases monotonically with the strand length $n$, reaching for large $n$ values much larger than the mean localization length $\lambda$ (dash-dotted line). From the tube model, we expect strands of length $n > N_e$ to be confined in a tube-like region of diameter $d \propto N_e^{1/2}$ \cite{rubinstein2003polymer}. After a brief transient, during which entanglements are not yet constraining the strands' dynamics, the strand monomers will fluctuate around their  equilibrium positions. For monomers that are part of short strands and are thus close to crosslinks, the fluctuations will be rather isotropic, whereas monomers which are in the middle of long strands will perform the largest excursions along the tube~\cite{lang2013monomer,lang2017relation}. Since the tube itself can be considered as a random walk, this motion is a ``random walk on a random walk", so that $\langle r_\text{mid}^2 \rangle \simeq d^2 (t/\tau_e)^{1/4}$, with $\tau_e \propto N_e^2$ the entanglement time. For a free chain, this regime will come to an end at the Rouse time $\tau_R \simeq \tau_e (n/N_e)^2$, when all the monomers start to move coherently and the chain diffuses along the tube, \textit{i.e.}, $\langle r_\text{mid}^2\rangle \propto t^{1/2}$. Our strands, however, are constrained at their ends by the crosslinks, so that instead the mean-squared displacement of the monomers reaches a pleateau with localization length~\cite{duering1994structure,lang2013monomer}

\begin{equation}
\lambda_\text{mid}(n) \simeq d \left(\frac n {N_e} \right)^{1/4} \qquad (n > N_e).
\end{equation}

\noindent The simulation data agree with the predicted scaling $\lambda_\text{mid}\propto n^{1/4}$, as shown in Fig.~\ref{fig:middle_plateau_400k}B, where we report $\rho \lambda_\text{mid}(n)$ as a function of $n$ for the systems of $4\cdot 10^5$ particles (here the $\rho$ prefactor accounts for the empirical observation that $\lambda \appropto \rho^{-1}$). One clearly sees that, whereas the small-$n$ behavior of $\rho \lambda_\text{mid}(n)$ depends on $f$ (and, to a lesser extent, on $c$), the large-$n$ behavior is basically independent of $f$ and $\rho \lambda_\text{mid}(n) \propto n^{1/4}$, confirming that the behavior of the longer strands is compatible with the tube model. %In addition, for small values of $n$, $\rho \lambda_\text{mid}(n)$ tends to a constant in agreement with Fig.~\ref{fig:msd_lambda}, confirming that the behavior of the average localization length is dominated by the shorter chains, which are the most abundant in our systems (see Fig.~\ref{fig:strand_length}B).

\begin{figure}
\centering
\includegraphics[width=0.46 \textwidth]{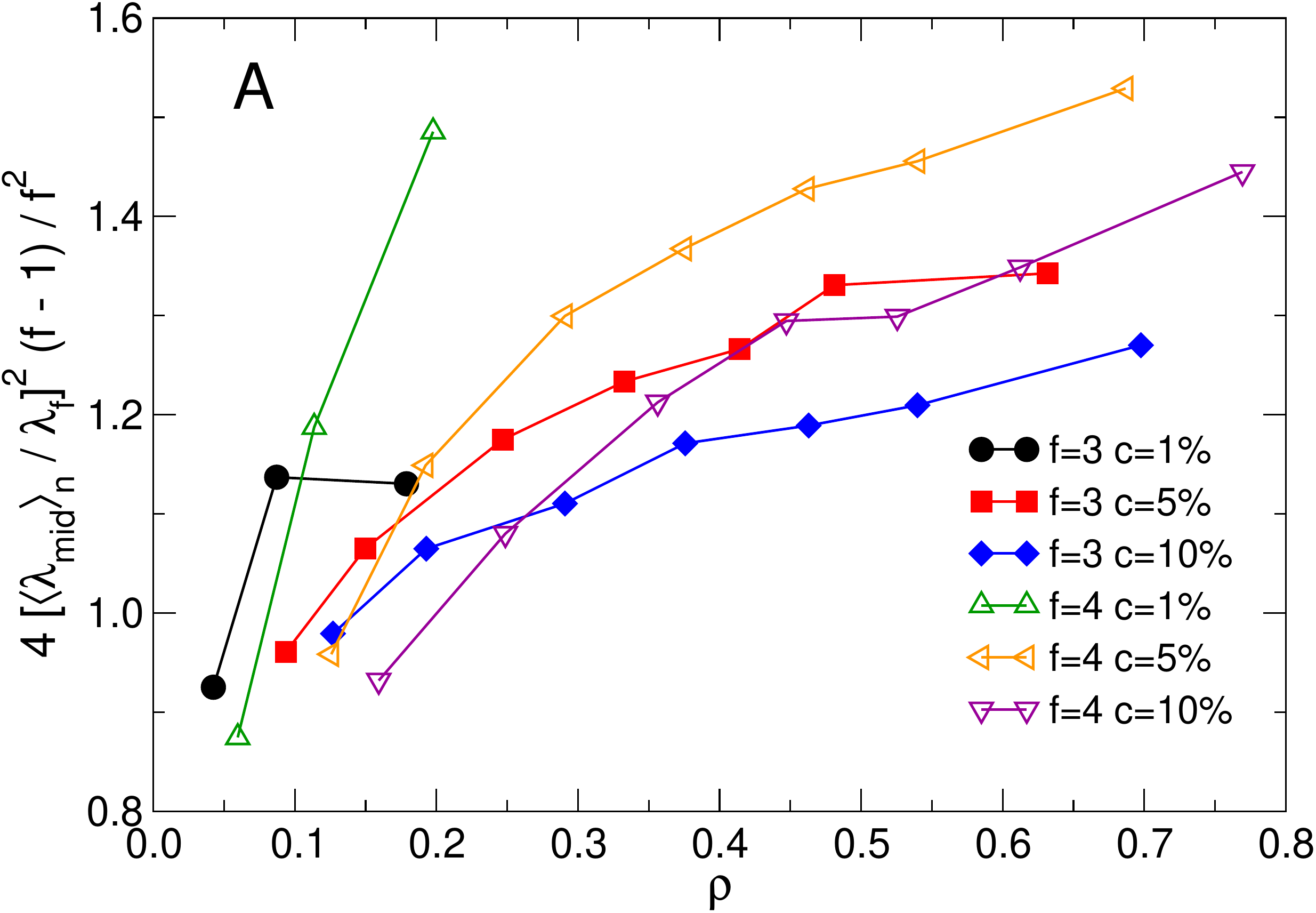}
\includegraphics[width=0.45 \textwidth]{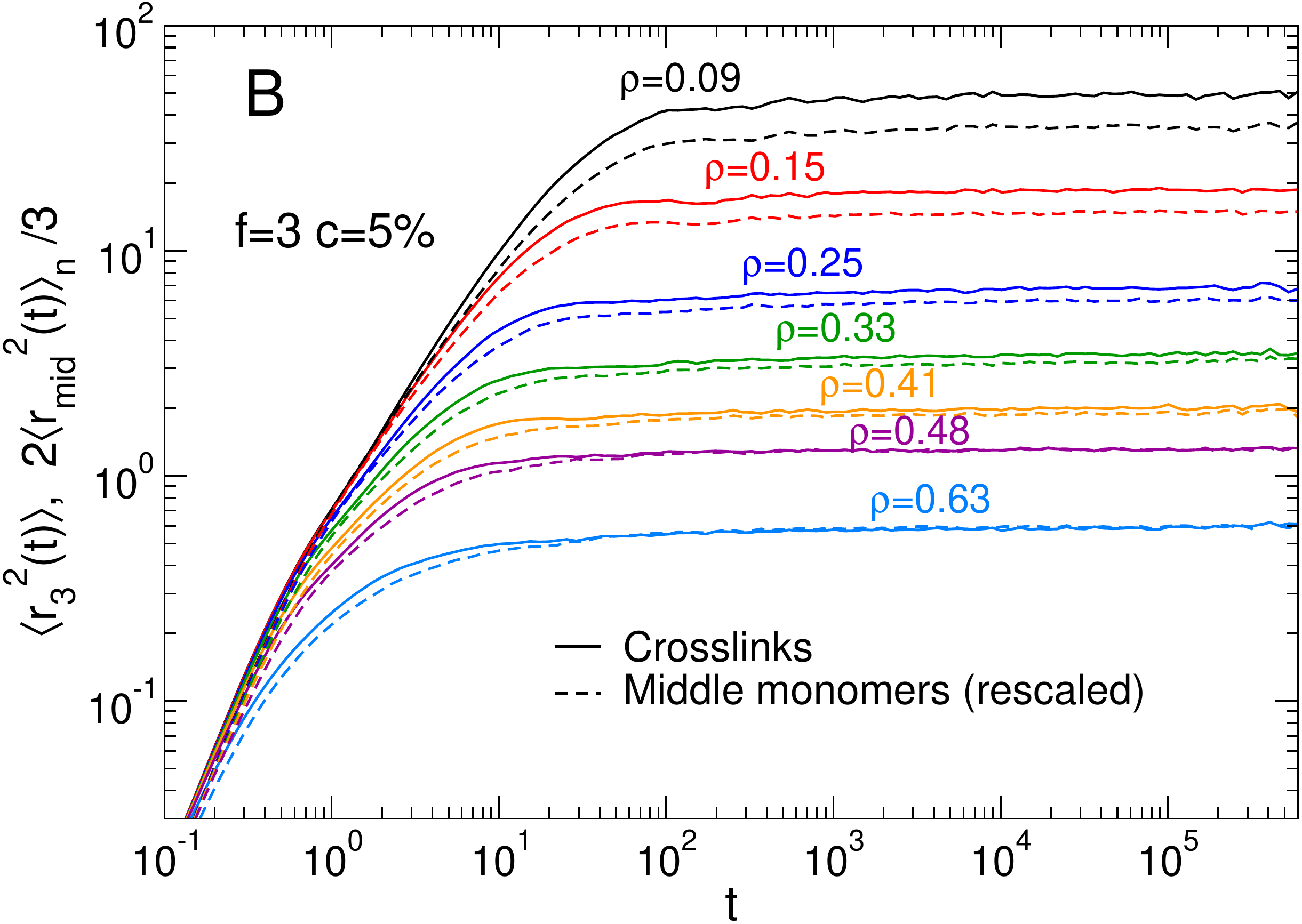}
\includegraphics[width=0.46 \textwidth]{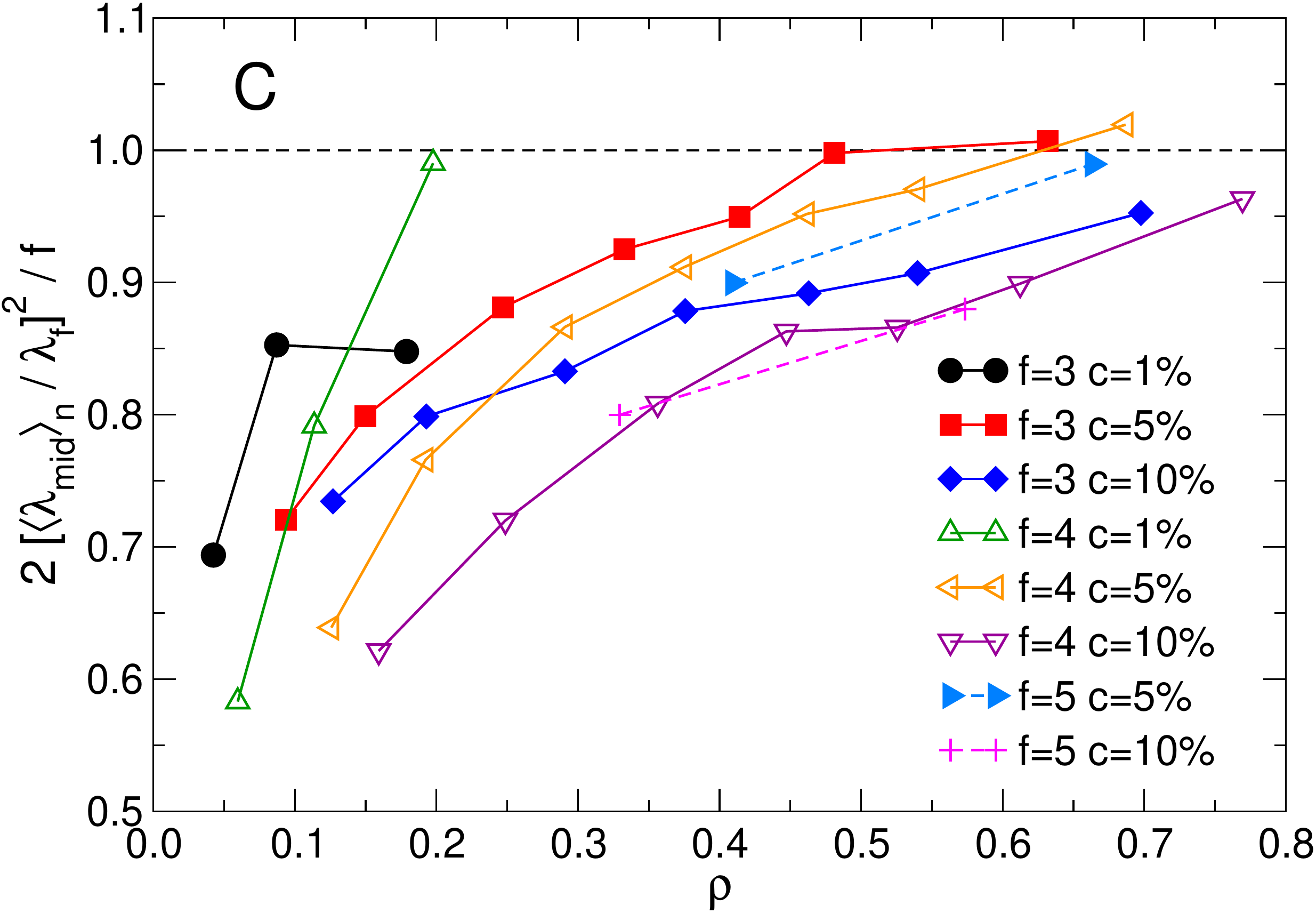}
\caption{(A) Ratio between the localization lengths of middle monomers and that of the crosslinks rescaled according to Eq.~\eqref{eq:localization_length_ratio_pnm}. (B) Comparison between the MSD of the crosslinks, $\langle r_f^2(t)\rangle$, and that of the middle monomers rescaled by Eq.~\eqref{eq:localization_length_ratio_high_rho}, $2\langle r_\text{mid}^2(t)\rangle / 3$, for systems with $f=3,c=5\%$ and different $\rho$. (C) Ratio between the localization lengths of middle monomers and that of the crosslinks rescaled according to Eq.~\eqref{eq:localization_length_ratio_high_rho}. The dashed line is the expected behaviour according to Eq.~\ref{eq:localization_length_ratio_high_rho}.}
\label{fig:msd_mid_cl}
\end{figure}

Fig.~\ref{fig:middle_plateau_400k}A also shows that, with the exception of the middle monomers of strands of length $n\leq 3$, the crosslinks are more constrained than the bivalent particles, as one may expect. In order to better compare the dynamics of these two types of particles, we compare the MSD of the crosslinks, $\langle r_f^2(t)\rangle$, to the average MSD of the middle monomers belonging to odd-length strands, $\langle r_\text{mid}^2(t)\rangle_n$ and their long-time limits $\langle \lambda_\text{mid}\rangle_n$ and $\lambda_f$, respectively (we recall that $\langle \cdot \rangle_n$ denotes an average over the configurations and over the strands). We start by computing the ratio $\langle \lambda_\text{mid}^2\rangle_n/\lambda_f^2$ in the framework of the phantom network model \cite{rubinstein2003polymer,kloczkowski1989chain,kloczkowski2002application} (PNM). In the PNM model, it is assumed that the network is composed by identical chains of size $N$ and each crosslink is connected to a fixed background through $f$ effective springs of constant $K_f = K (f - 2)/(f - 1)$, where $K = 3 k_B T / b^2 N$ is the ideal entropic spring constant of each chain and $b$ is the  Kuhn length. From the equipartition theorem \cite{huang1963statistical}, one obtains $\lambda_f^2(N) = {3 k_B T}/(f K_f) = b^2N (f-1)/[f(f-2)]$ which, averaged over the exponential chain-size distribution that characterizes our networks (see Fig.~\ref{fig:strand_length}B), becomes

\begin{equation}
\lambda_f^2 = b^2N_s \frac{f-1}{f(f-2)}.
\label{eq:localization_length_f}
\end{equation}

\noindent
Under the same approximation, the middle monomer of a strand of size $N_s$ is connected to two crosslinks by chains of length $N_s / 2$, which behave as entropic springs of constant $2 K = 6 k_B T / b^2 N_s$. In turn, each crosslink is connected to the fixed background through $(f - 1)$ springs of constant $K_f$. Therefore, recalling that the effective constant of a set of springs is the sum of the constants if the springs are connected in parallel, and the reciprocal of the sum of the reciprocal constants if the springs are connected in series, a middle monomer is connected to the fixed background through two springs of constant $2K(f-2)/2$ or, equivalently, through a single spring of constant $4 K (f - 2) / f$. As a result, the PNM model predicts for the localization length of middle monomers (averaged over the exponential chain-size distribution as done above):

\begin{equation}
\langle \lambda_\text{mid}^2 \rangle_n = b^2 N_s\frac{f}{4(f - 2)}.
\label{eq:localization_length_mid}
\end{equation}

\noindent
Dividing Eq.~\eqref{eq:localization_length_mid} by Eq.~\eqref{eq:localization_length_f} yields

\begin{equation}
\frac{\langle \lambda_\text{mid}^2\rangle_n}{\lambda_f^2} = \frac{f^2}{4(f - 1)}.
\label{eq:localization_length_ratio_pnm}
\end{equation}

\noindent
Figure~\ref{fig:msd_mid_cl}A shows that rescaling the numerical data for $\langle \lambda_\text{mid}^2\rangle_n/\lambda_f$ by Eq.~\eqref{eq:localization_length_ratio_pnm} yields values that at low density are close to $1$, and hence to the PNM prediction, for all the systems. However, all curves retain a dependence on $c$ and a significant dependence on the density as $\rho \to 0$ that are not captured by the simplistic PNM approach.

As the density increases, excluded volume effects and the non-crossability of the strands become important, leading to a monotonic increase of $\langle \lambda_\text{mid}^2\rangle_n/\lambda_f^2$. We attempt to rationalize this behavior by hypothesizing that, at high density, the large number of topological constraints increases the effective spring constant through which a particle is connected to the fixed background to a value $K_\text{eff}$ that is the same for every particle, be it a monomer or a crosslink. Under this assumption, the localization length of a particle is given by

\begin{equation}
\label{eq:localization_length_N}
\lambda_\mathcal{N}^2 = \frac{3 k_B T}{\mathcal{N} K_\text{eff}},
\end{equation}

\noindent
where $\mathcal{N}$ is the effective number of chains to which it is connected. Therefore, with this \textit{ansatz}, the high-density limit of the ratio between the localization lengths is simply given by

\begin{equation}
\frac{\langle \lambda_\text{mid}^2\rangle_n}{\lambda_f^2} = \frac{f}{2}.
\label{eq:localization_length_ratio_high_rho}
\end{equation}

\noindent
Fig.~\ref{fig:msd_mid_cl}B shows that the MSD of the middle monomers multiplied by a factor $2/f$ overlaps well with the MSD of the crosslinks at high density, with the agreement getting worse as $\rho$ decreases. We also note that values in agreement with Eq.~\eqref{eq:localization_length_ratio_high_rho} have been found in other simulation studies of similar systems \cite{russo2009reversible, kenkare1998discontinuous}. Finally, in Fig.~\ref{fig:msd_mid_cl}C we rescale $\langle \lambda_\text{mid}^2\rangle_n/\lambda_f^2$ by $f / 2$ for all the investigated systems, finding that all curves tend towards $1$ as $\rho$ increases, supporting the hypotheses underlying Eq.~\eqref{eq:localization_length_ratio_high_rho}. To test the more general applicability of Eq.~\eqref{eq:localization_length_ratio_high_rho} for different values of $f$, we also include in Fig.~\ref{fig:msd_mid_cl}C results for networks with pentavalent ($f=5$) crosslinks, finding results in agreement with those observed for $f=3$ and $4$. We note that in both the low- and high-density limit the theoretical arguments we put forward are not sufficient to explain the $c$-dependence exhibited by the numerical data.

%%%%%%%%%%%%%%%%%%%%%%%%%%%%%%%%%%%%%%%%%%%%%%%%%%%%%%
\subsection{Poisson ratio and elastic moduli}
\label{subsec:elastic_moduli}
%%%%%%%%%%%%%%%%%%%%%%%%%%%%%%%%%%%%%%%%%%%%%%%%%%%%%%

\begin{figure}
\centering
\includegraphics[width=0.45 \textwidth]{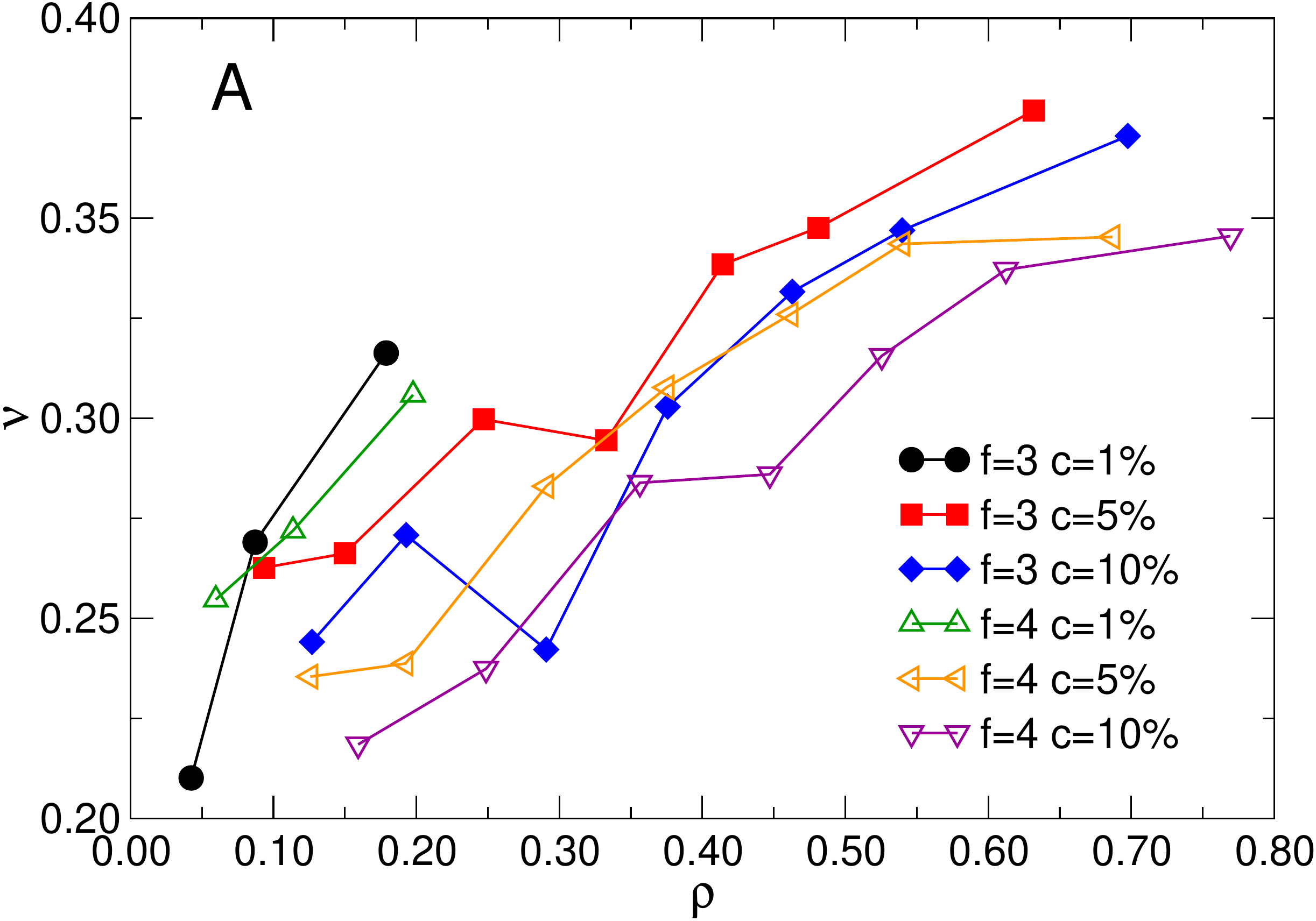}
\includegraphics[width=0.45 \textwidth]{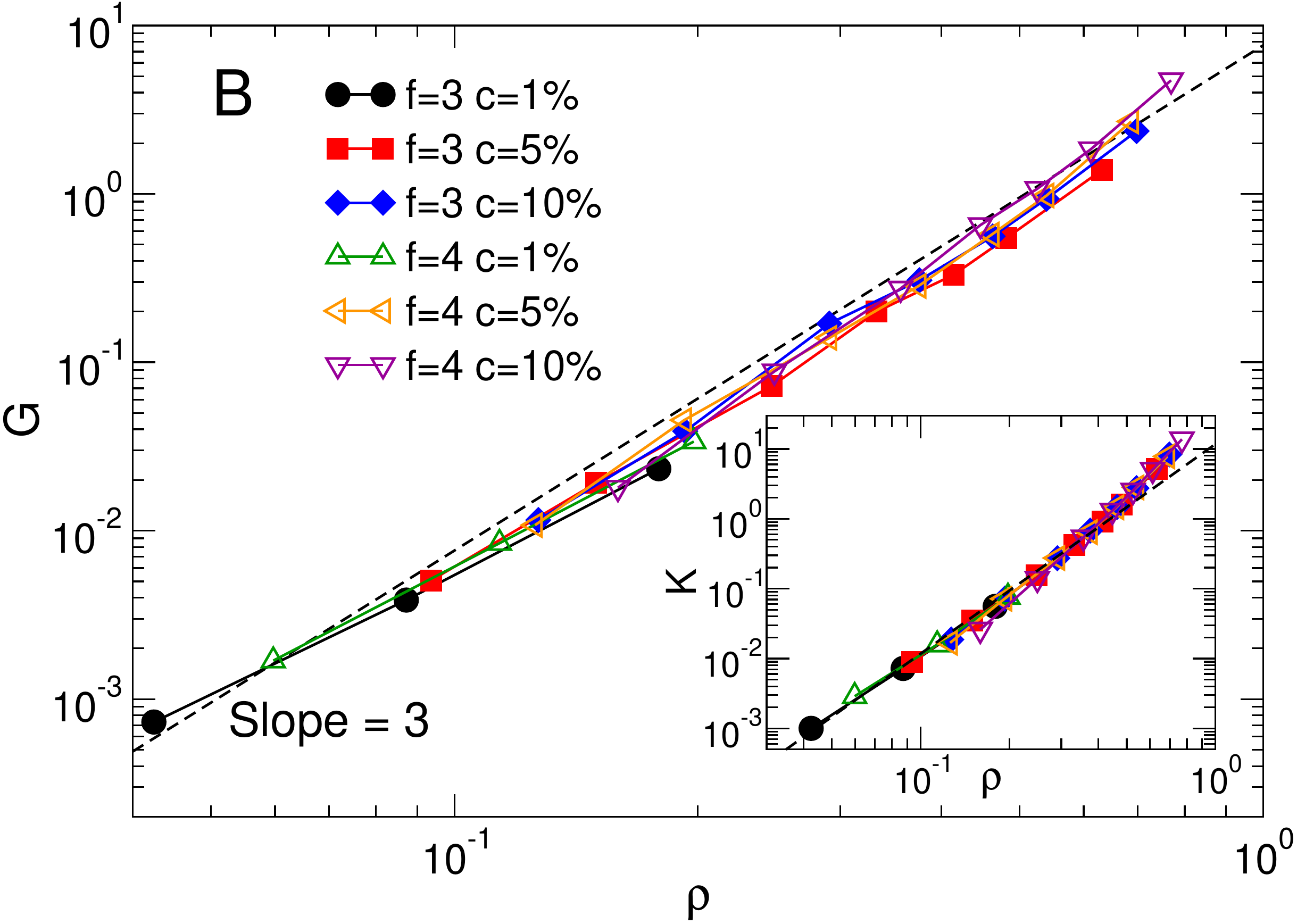}
\caption{(A) The Poisson ratio $\nu_P$ and (B) the shear modulus $G$ of the network as functions of monomer density $\rho$ for different values of $f$ and $c$. \textit{Inset}: bulk modulus $K$. The dashed line corresponds to a $\rho^3$ behaviour.}
\label{fig:elasticity}
\end{figure}

The elasticity of polymer networks is usually characterised by the shear modulus, $G$ \cite{rubinstein2003polymer}. Since polymer gels are, in general, compressible, they possess a finite bulk modulus, $K$, and a Poisson ratio $\nu_P$ that is smaller than $1/2$ \cite{pritchard2013swelling}. Here we use the method of Ref.  \cite{rovigatti2019connecting} to estimate $G$ and $K$ from the fluctuations of the box size in the three spatial dimensions, making it possible to compute $\nu$ through the relation

\begin{equation}
\nu_P = \frac{3K - 2G}{6K + 2G}.
\label{eq:nu}
\end{equation}

\noindent
Using theoretical arguments building on the Flory-Huggins theory and the elasticity of the phantom network it has been shown that the Poisson ratio of gels goes from $\nu_P \simeq 0.25$ for swollen samples to $\nu_P \simeq 1/2$ for rubbers and melts, which are essentially incompressible \cite{boon2017swelling,pritchard2013swelling}. Figure~\ref{fig:elasticity}A shows that the polymer networks we investigate have a Poisson ratio that is between $\simeq 0.20$ and $\simeq 0.25$ in the maximally swollen state and then, within the numerical noise, increases monotonically as density increases, therefore comparing favourably, at least on a semi-quantitative level, with the theoretical figures reported above. Interestingly, the dependence on $c$ and $f$ is rather weak and mostly masked by the statistical uncertainties.

The $\rho$-dependence of the shear modulus $G$, shown in Fig.~\ref{fig:elasticity}B, is much stronger than that of $\nu$. Indeed, the relation $G \propto \rho^3$ approximately holds in the whole density range, but in particular at high density. Unraveling the different contributions that make up the shear modulus of a polymer network is a long-standing open problem in polymer physics \cite{zhong2016quantifying,hoshino2018network,matsuda2019fabrication,lin2019revisiting,gula2020computational, lang2022reference}, especially when short chains are present, as it is the case here. Indeed, previous simulations of polydisperse and disordered phantom networks (\textit{i.e.}, networks where excluded volume interactions are turned off), have established that classical elasticity theory \cite{treloar2005physics} fails to describe the elastic behavior of the system when short chains are abundant~\cite{sorichetti2021effect}. Moreover, the structures of the networks varies with density, a fact that prevents using classical elasticity theory to extrapolate the shear modulus $G$ from small to large densities, or \textit{vice versa}. As a result, in these cases both the phantom and affine network models fail to capture even the qualitative behaviour of the network elastic moduli. 
In light of this, we can nonetheless attempt rationalize the $\rho^3$ dependence of the shear modulus by attempting to connect it with the behavior of the localization length as follows: We start by recalling that in the simulated systems, the localization length of the crosslinks behaves, similarly to the total localization length (Fig.~\ref{fig:msd_lambda}B), as $\lambda_f \appropto \rho^{-1}$ (see Fig.~\ref{fig:lambda_all} in the Appendix). In line with the results shown in Sec.~\ref{sec:lambda} (in particular Eq.~\eqref{eq:localization_length_N}), we then assume that the squared localization length of the particles is inversely proportional to an effective spring constant that depends on the density, $\lambda_f^2 \propto K_\text{el}^{-1}(\rho)$, so that $K_\text{el}(\rho) \appropto \rho^2$. By interpreting $K_\text{el}$ as the entropic spring constant of a chain of length $N_\text{el} \propto K_\text{el}^{-1}$ and assuming that the shear modulus is proportional to the density $\rho/N_\text{el}$ of these effective chains \cite{rubinstein2003polymer}, we find

\begin{equation}
G(\rho) \propto \frac{\rho}{N_\text{el}} \appropto \rho^3,
\label{eq:G}
\end{equation}

\noindent
in agreement with the observed $\rho$-dependence and the deviations at small $\rho$ observed in Fig.~\ref{fig:msd_mid_cl}C. In the inset of Fig.~\ref{fig:elasticity}B, we also report the bulk modulus $K$, which behaves roughly as $G$. This is expected, as $\nu_P$ only changes slightly in the density range studied, and from Eq.~\eqref{eq:nu} one has $K/G = 2(1+\nu_P)/[3(1-2\nu_P)]$. We recall that scaling arguments \cite{rubinstein2003polymer} predict $K \propto \rho^{3/(3-d_f)}$, $d_f$ being the fractal dimension of the network. Recent simulations of compact nanogel particles \cite{chremos2022molecular} and bottle-brush polyelectrolytes \cite{horkay2021comparative} yielded $d_f=2$. This fractal dimension is consistent with that theoretically predicted for randomly branched polymers with excluded volume interactions \cite{parisi1979hausdorff}, and further implies $K \propto \rho^3$, which is indeed the density scaling observed here. We note that a different behaviour, compatible with $d_f=1.7$ (the fractal dimension of linear chains in good solvent), was recently reported for tetra-PEG gels \cite{yasuda2020universal}. Altogether, the results indicate that the nature of the system and of its assembly process can affect the observed fractal dimension and, by extension, the scaling of the elastic modulus too.
%%%%%%%%%%%%%%%%%%%%%%%%%%%%%%%%%%%%%%%%%%%%%%%%%%%%%%%
%%%%%%%%%%%%%%%%%%%%%%%%%%%%%%%%%%%%%%%%%%%%%%%%%%%%%%%
\section{Discussion and conclusions}
\label{sec:conclusions}
%%%%%%%%%%%%%%%%%%%%%%%%%%%%%%%%%%%%%%%%%%%%%%%%%%%%%%%
%%%%%%%%%%%%%%%%%%%%%%%%%%%%%%%%%%%%%%%%%%%%%%%%%%%%%%%

Up to the present day most simulation studies of polymer networks considered structures that are either ordered, monodisperse (with a unique strand length) or both~\cite{sonnenburg1990molecular,everaers1995test,escobedo1996monte,everaers1996topological, escobedo1997simulation, everaers1999entanglement}. Even in the case of networks produced through random crosslinking of precursor chains, it is very difficult, due to the slow chain dynamics, to reach a fully-bonded state in which dangling ends are absent\cite{grest1992kinetics,duering1994structure,putz2000self,gilra2000monte,kenkare1998discontinuous, lang2020analysis, lang2022reference}. By contrast, here we used a method to generate disordered, polydisperse networks which are almost defect-free from the assembly of bivalent and $f$-valent (crosslinks) patchy particles~\cite{gnan2017silico,sorichetti2021effect,ninarello2022onset}. In the present study, all the networks satisfied more than $99.8\%$ of all possible bonds and no more than $4\%$ of the monomers were part of dangling ends, but it is in principle possible, with a larger computational effort, to obtain disordered fully-bonded networks ($100\%$ of bonds formed) with no dangling ends. Moreover, the assembly process is an equilibrium one: as a result, the final structure of the network depends only on the percentage $c$ of crosslinks, on their valence $f$ and on the assembly density $\rho_\text{init}$. Another interesting property of these networks is that the distribution $m_n$ of strand lengths is independent of density, and only depends on $f$ and $c$. However, we showed that the number of topological kinks depends rather strongly on density, and increases approximately as a power law: This makes it possible to tune the entanglement length by changing the assembly density, while keeping the average strand length $N_s$ constant. We further note that the same model considered here can be used to study the influence that $f$ and $c$ have on the local network structure, \textit{e.g.}, ring statistics. Likewise, the model can be used to investigate how short-range attractions that mimic the solvent quality during the assembly will impact the final network.

By analysing the dynamical properties of the networks on the level of the strands, we found that the dynamics of long strands is qualitatively consistent with the tube model. Moreover, we rationalised the behavior of the ratio between the localization length of the crosslinks, $\lim_{t \to \infty} \langle r^2_f(t)\rangle$, and the localization length of the middle monomers of the strands, $\lim_{t \to \infty} \langle r^2_\text{mid}(t)\rangle$, finding a crossover between a phantom network model-like behavior at low density and a mean-field-like regime at high density where each particle can be described as being connected \textit{via} an effective spring.
Finally, we discussed the elasticity of the networks, showing that the values of the Poisson ratio we observe are in line with experimental values, and that the peculiar $\rho$-dependence of the network shear modulus we observe, $G \propto \rho^3$, can be interpreted in light of the behavior of the localization length of the particles discussed above. We mention that the study of how the localization length and its relation to the shear modulus change when the network is subject to deformations  is a promising direction for future work. Overall, the results presented here show that the assembly method we have used yields polymer networks that display realistic properties, and thus can be used to model interesting phenomena where the polydisperse and disordered nature of the networks become important, such as polymer-nanoparticle composites~\cite{sorichetti2021dynamics} or double networks~\cite{ducrot2014toughening}. However, the peculiar scaling (or apparent scaling) behaviour we found here cannot be straightforwardly understood by using classic polymer theories in view of the presence of short chains that are abundant in the systems investigated~\cite{sorichetti2021effect}. Therefore, the results we have shown here should be taken as motivation for future theoretical efforts attempting to describe the behaviour of realistic disordered polymer networks.

%%%%%%%%%%%%%%%%%%%%%%%%%%%%%%%%%%%%%%%%%%%
%%%%%%%%%%%%%%%%%%%%%%%%%%%%%%%%%%%%%%%%%%%
\section*{Acknowledgements}

We thank Michael Lang for helpful discussions. We acknowledge financial support from the European Research Council (ERC Consolidator Grant 681597, MIMIC) and from LabEx NUMEV
(ANR-10-LABX-20) funded by the ‘‘Investissements d'Avenir’’
French Government program, managed by the French National Research Agency
(ANR). W.K.~is a senior member of the Institut Universitaire de France. 

\section*{Data Availability Statement}

The data that support the findings of this study are available from the corresponding author upon request.

\section*{Author Declarations}

The authors have no conflicts to disclose

%%%%%%%%%%%%%%%%%%%%%%%%%%%%%%%%%%%%%%%%%%%
%%%%%%%%%%%%%%%%%%%%%%%%%%%%%%%%%%%%%%%%%%%

%%%%%%%%%%%%%%%%%%%%%%%%%%%%%%%%%%%%%%%%%%%
%%%%%%%%%%%%%%%%%%%%%%%%%%%%%%%%%%%%%%%%%%%
\section*{Appendix} 
%%%%%%%%%%%%%%%%%%%%%%%%%%%%%%%%%%%%%%%%%%%
%%%%%%%%%%%%%%%%%%%%%%%%%%%%%%%%%%%%%%%%%%%

\setcounter{section}{0}
\setcounter{equation}{0}

\renewcommand{\thesection}{A\Roman{section}}
\renewcommand{\theequation}{A\arabic{equation}}

%%%%%%%%%%%%%%%%%%%%%%%%%%%%%%%%%%%%%%%%%%%
\section{Additional details on the simulated systems}
\label{app:details}
%%%%%%%%%%%%%%%%%%%%%%%%%%%%%%%%%%%%%%%%%%%

In Tabs.~\ref{tab:f3} and~\ref{tab:f4} we report the details of the simulated networks, respectively with trivalent ($f=3$) and tetravalent ($f=4$) crosslinks. $M_{tot}$: Total number of particles ($M_2+M_f)$. $M_4$: Number of tetravalent particles. $M_3$: Number of trivalent particles. $M_2$: Number of bivalent particles. $c\equiv (M_4+M_3)/M_\text{tot}$: crosslink concentration. $F$: mean valence, Eq.~\eqref{eq:mean_valence}. $\rho\equiv M_\text{tot}/\langle V \rangle$: average density ($\langle V \rangle$ = configuration-averaged volume). All the values are averaged over two independent realizations of the system.

\begin{table*}
\centering
$\begin{array}{@{\hspace{2em}} c @{\hspace{2em}} c @{\hspace{2em}} c @{\hspace{2em}} c @{\hspace{2em}} c @{\hspace{2em}} c @{\hspace{2em}} }
\hline
M_\text{tot}  & M_3 & M_2 & c & F & \rho\\
\hline
48773.0 & 4845.0 & 43928.0 & 0.0993 & 2.099 & 0.1270\\ 
49388.5 & 4907.0 & 44481.5 & 0.0994 & 2.099 & 0.1929\\ 
49636.0 & 4940.0 & 44696.0 & 0.0995 & 2.100 & 0.2908\\ 
49731.5 & 4953.5 & 44778.0 & 0.0996 & 2.100 & 0.3758\\ 
49803.0 & 4966.0 & 44837.0 & 0.0997 & 2.100 & 0.4630\\ 
49797.5 & 4968.5 & 44829.0 & 0.0998 & 2.100 & 0.5398\\ 
49714.0 & 4961.5 & 44752.5 & 0.0998 & 2.100 & 0.6976\\ 
\hline
47023.0 & 2346.0 & 44677.0 & 0.0499 & 2.050 & 0.0936\\ 
48508.0 & 2394.0 & 46114.0 & 0.0494 & 2.049 & 0.1498\\ 
48976.0 & 2434.0 & 46542.0 & 0.0497 & 2.050 & 0.2469\\ 
49243.5 & 2454.0 & 46789.5 & 0.0498 & 2.050 & 0.3327\\ 
49439.0 & 2465.0 & 46974.0 & 0.0499 & 2.050 & 0.4141\\ 
49469.0 & 2465.0 & 47004.0 & 0.0498 & 2.050 & 0.4812\\ 
49231.5 & 2457.0 & 46774.5 & 0.0499 & 2.050 & 0.6317\\ 
\hline
36368.5 &352.0&36016.5& 0.0097  &2.010  &0.0425\\
41791.0 &405.0  &41386.0&   0.0097  &2.010  &0.0872\\
46047.5 &441.0& 45606.5&    0.0096& 2.010   &0.1789\\
\hline
\end{array}$
\caption{Properties of the networks with trivalent ($f=3$) crosslinks.}
\label{tab:f3}
\end{table*}

\begin{table*}
\centering
$\begin{array}{@{\hspace{1.5em}} c @{\hspace{1.5em}} c @{\hspace{1.5em}} c @{\hspace{1.5em}} c @{\hspace{1.5em}} c @{\hspace{1.5em}} c @{\hspace{1.5em}} c @{\hspace{1.5em}} }
\hline
M_\text{tot} & M_4 & M_3 & M_2 & c & F & \rho\\
\hline
48952.0 & 4821.5 & 176.0 & 43954.5 & 0.1021 & 2.201 & 0.1593\\ 
49568.5 & 4901.5 & 98.5 & 44569.0 & 0.1009 & 2.200 & 0.2487\\ 
49748.0 & 4924.5 & 75.0 & 44748.5 & 0.1005 & 2.199 & 0.3563\\ 
49792.5 & 4948.0 & 52.0 & 44792.5 & 0.1004 & 2.200 & 0.4472\\ 
49804.5 & 4958.0 & 42.0 & 44804.5 & 0.1004 & 2.200 & 0.5255\\ 
49888.0 & 4970.0 & 29.0 & 44889.0 & 0.1002 & 2.200 & 0.6123\\ 
49718.5 & 4998.5 & 71.5 & 44720.0 & 0.1005 & 2.200 & 0.7694\\ 
\hline
47464.5 & 2319.5 & 163.0 & 44982.0 & 0.0523 & 2.101 & 0.1258\\ 
48946.5 & 2408.5 & 91.0 & 46447.0 & 0.0511 & 2.100 & 0.1924\\ 
49469.0 & 2441 & 58.5 & 46969.5 & 0.0505 & 2.100 & 0.2908\\ 
49439.0 & 2450.5 & 49.0 & 46939.5 & 0.0506 & 2.100 & 0.3750\\ 
49620.0 & 2462.0 & 37.5 & 47120.0 & 0.0504 & 2.100 & 0.4617\\ 
49700.5 & 2466.0 & 34.0 & 47200.5 & 0.0503 & 2.100 & 0.5399\\ 
49494.5 & 2500.0 & 55.5 & 46994.5 & 0.0505 & 2.100 & 0.6868\\ 
\hline
39123.5&    373.5&  105.0&  38645.0&    0.0122& 2.022 & 0.0597\\
44301.5&    419.0&  72.0    &43810.5&   0.0111& 2.021 & 0.1137\\
46274.5&    438.0&  58.0    &45778.5&   0.0107& 2.020 & 0.1977\\
\hline
\end{array}$
\caption{Properties of the networks with tetravalent ($f=4$) crosslinks.}
\label{tab:f4}
\end{table*}

\begin{figure}[t]
\centering
\includegraphics[width=0.45 \textwidth]{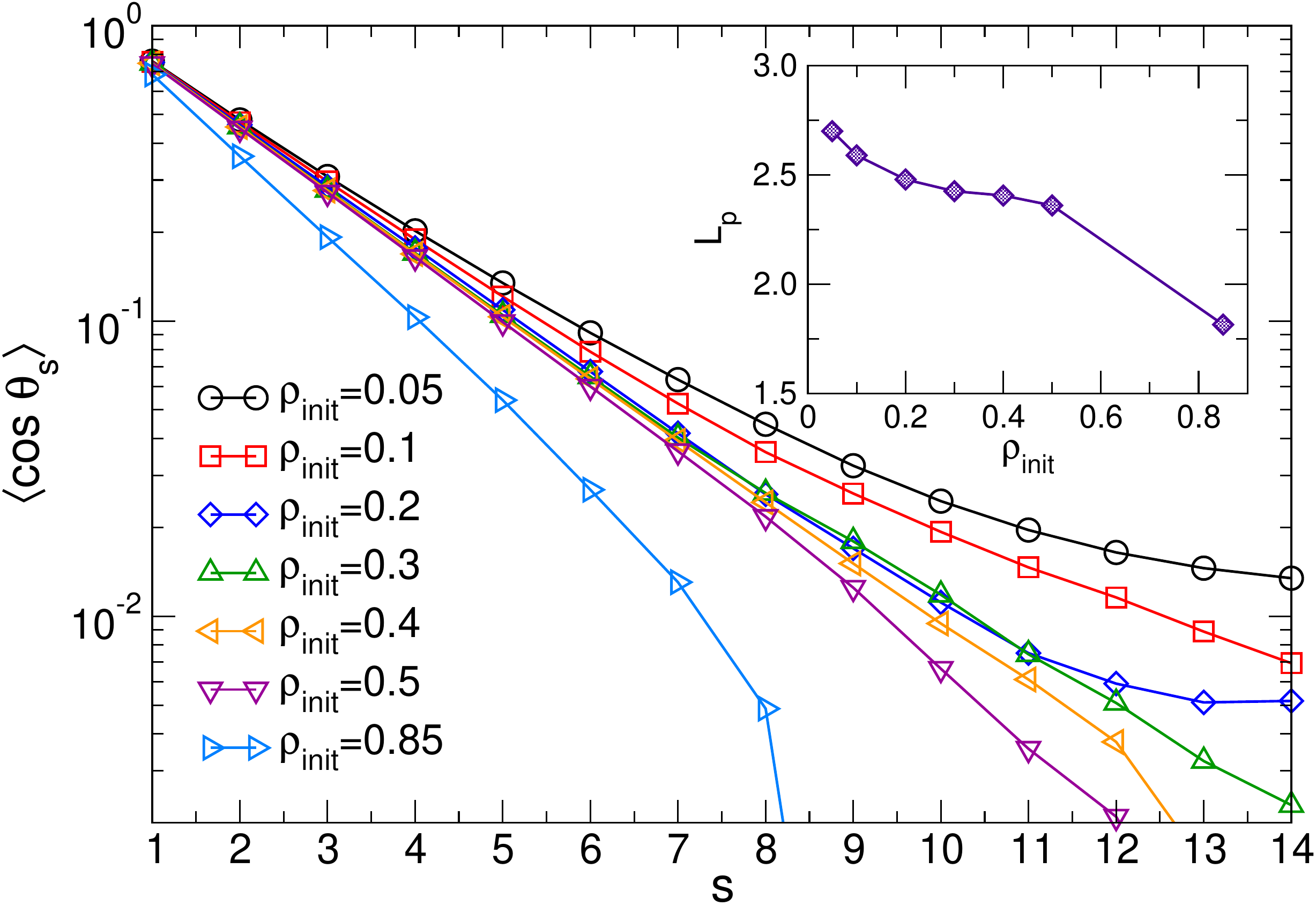}
\caption{Bond-bond orientational correlation function for systems of di-valent patchy particles assembled at different densities $\rho_\text{init}$. Inset: persistence length $L_p$ obtained by fitting the first 5 points of each curve to Eq.~\ref{eq:bond_corr_exp}.} 
\label{fig:costheta}
\end{figure}

To characterize the persistence length of the patchy particle model, we started from a monomer solution and used the step-growth polymerization \cite{zhang2022dispersity}
to assemble chains of patchy particles. For the latter, we next calculated the bond-bond orientational correlation function $\langle \cos(\theta_s)\rangle$, defined as \cite{wittmer2004long}

\begin{equation}
\langle \cos(\theta_s) \rangle \equiv  \left\langle \frac{\mathbf b_k \cdot \mathbf b_{k+s}}{|\mathbf b_k| \ |\mathbf b_{k+s}|} \right\rangle,
\label{eq:bond_corr_def}
\end{equation}

\noindent
where $\mathbf b_k \equiv\mathbf r_{k+1} - \mathbf r_k$ is the $k$-th bond vector, $\langle \rangle$ denote ensemble averages taken over all bond vectors separated by a chemical distance $s$ (\textit{i.e.} over all those pairs of bond vectors belonging to the same chain and with sequence separation $s$). The results are shown in Figure~\ref{fig:costheta}. At all considered values of $\rho_{\rm init}$, the bond-bond correlation function has an approximate exponential decay for $ 1 \le s \le 6$, which we exploited to extract the persistence length, $L_p$, from the best fit of the data to the functional form\\

\begin{equation}
\langle \cos(\theta_s)\rangle \propto e^{-sb/L_p},
\label{eq:bond_corr_exp}
\end{equation}

\noindent where $b$ is the bond length and $L_p$ is the persistence length. The bond length depends weakly on $\rho_\text{init}$: it takes the value $b=1.2$ for $\rho_\text{init}=0.05$ and decreases by only $\simeq 5\%$ when $\rho_\text{init}$ is increased to $0.85$.
The best fit procedure yields a persistence length that decreases monotonically with density, ranging between $2.7$ for $\rho_\text{init}=0.05$ and $1.8$ for $\rho_\text{init}=0.85$, as shown in the inset of Fig.~\ref{fig:costheta}.

%%%%%%%%%%%%%%%%%%%%%%%%%%%%%%%%%%%%%%%%%%%
\section{Localization length}
%%%%%%%%%%%%%%%%%%%%%%%%%%%%%%%%%%%%%%%%%%%

\begin{figure}
\centering
\includegraphics[width=0.45 \textwidth]{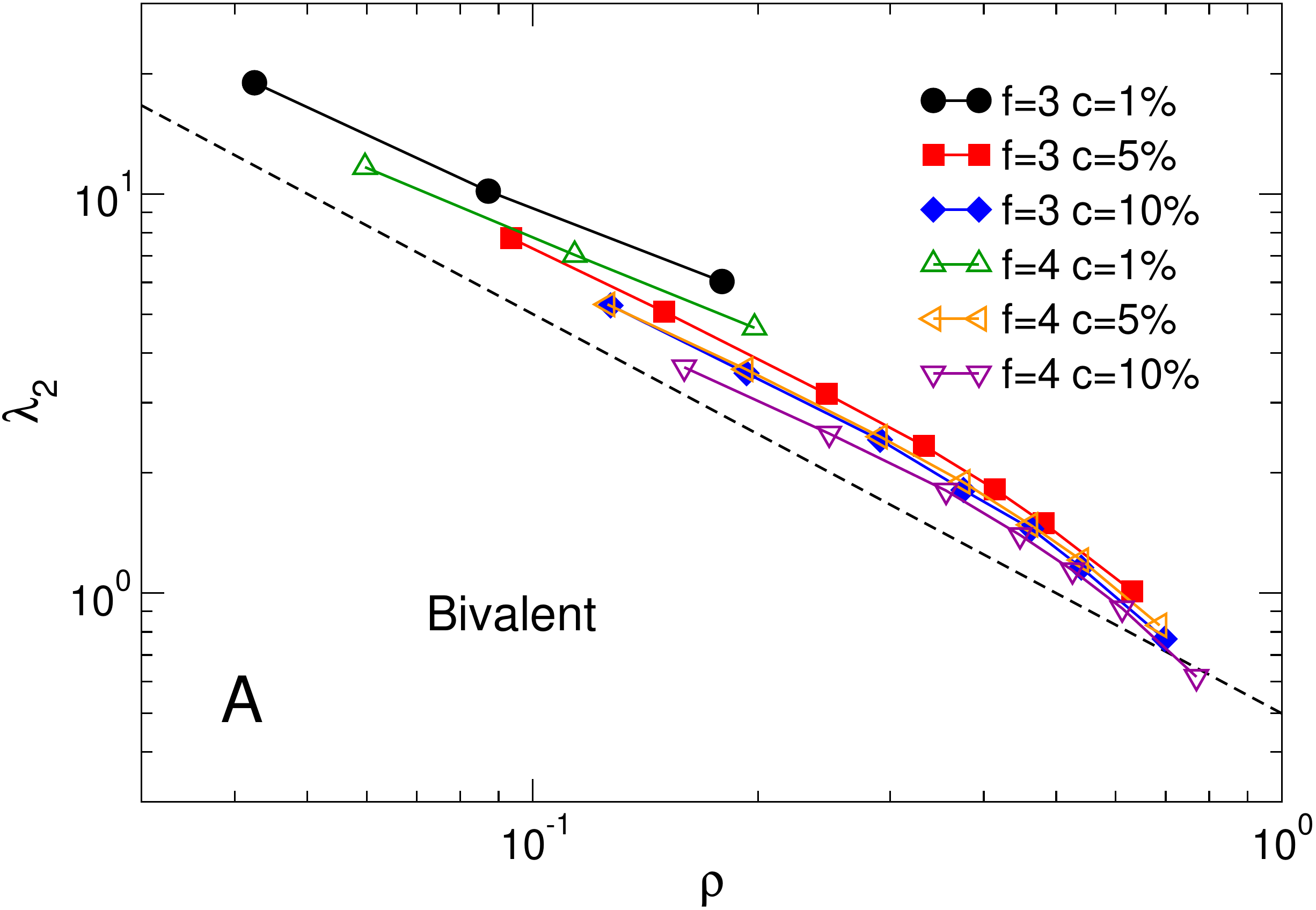}
\includegraphics[width=0.45 \textwidth]{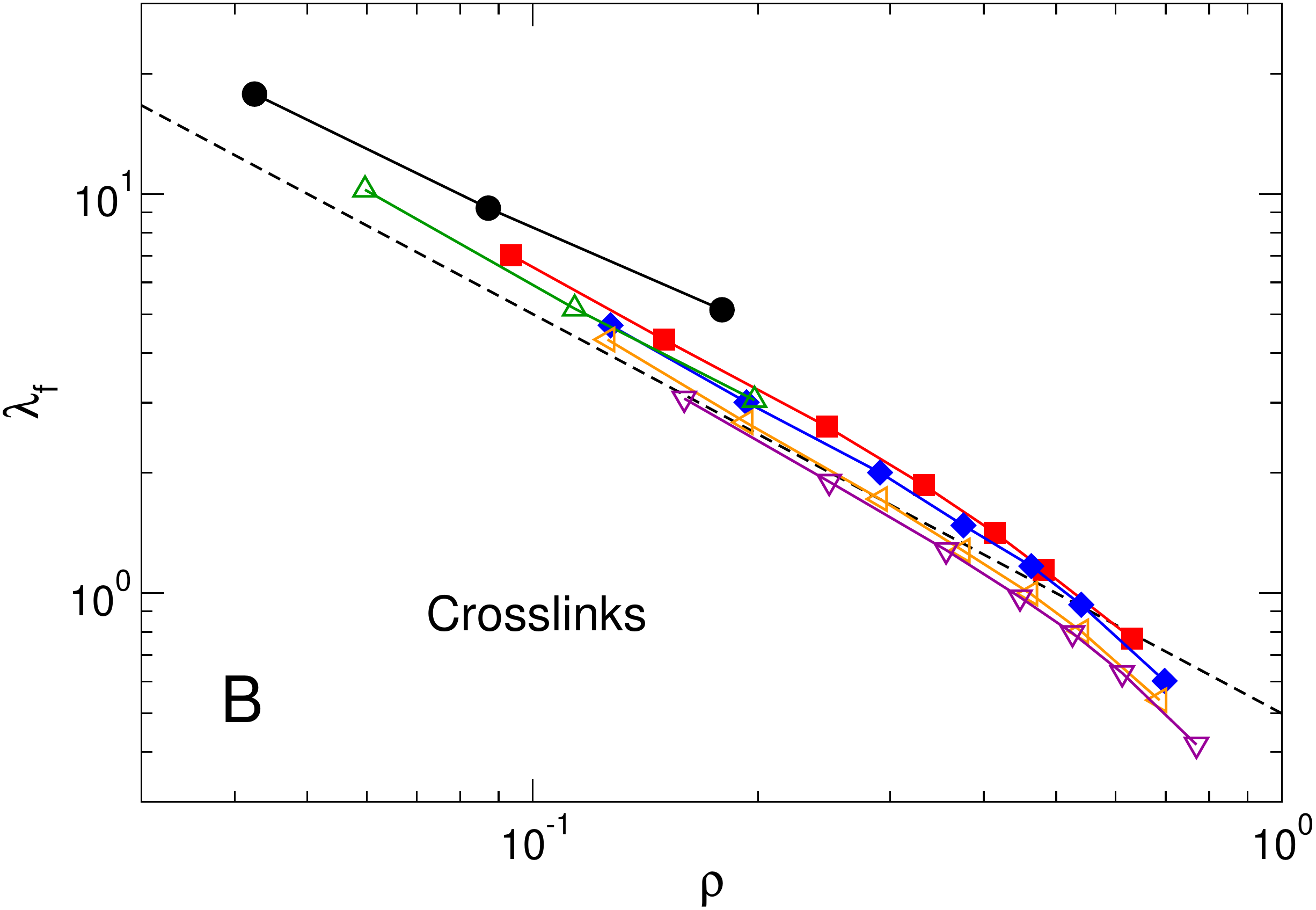}
\includegraphics[width=0.45 \textwidth]{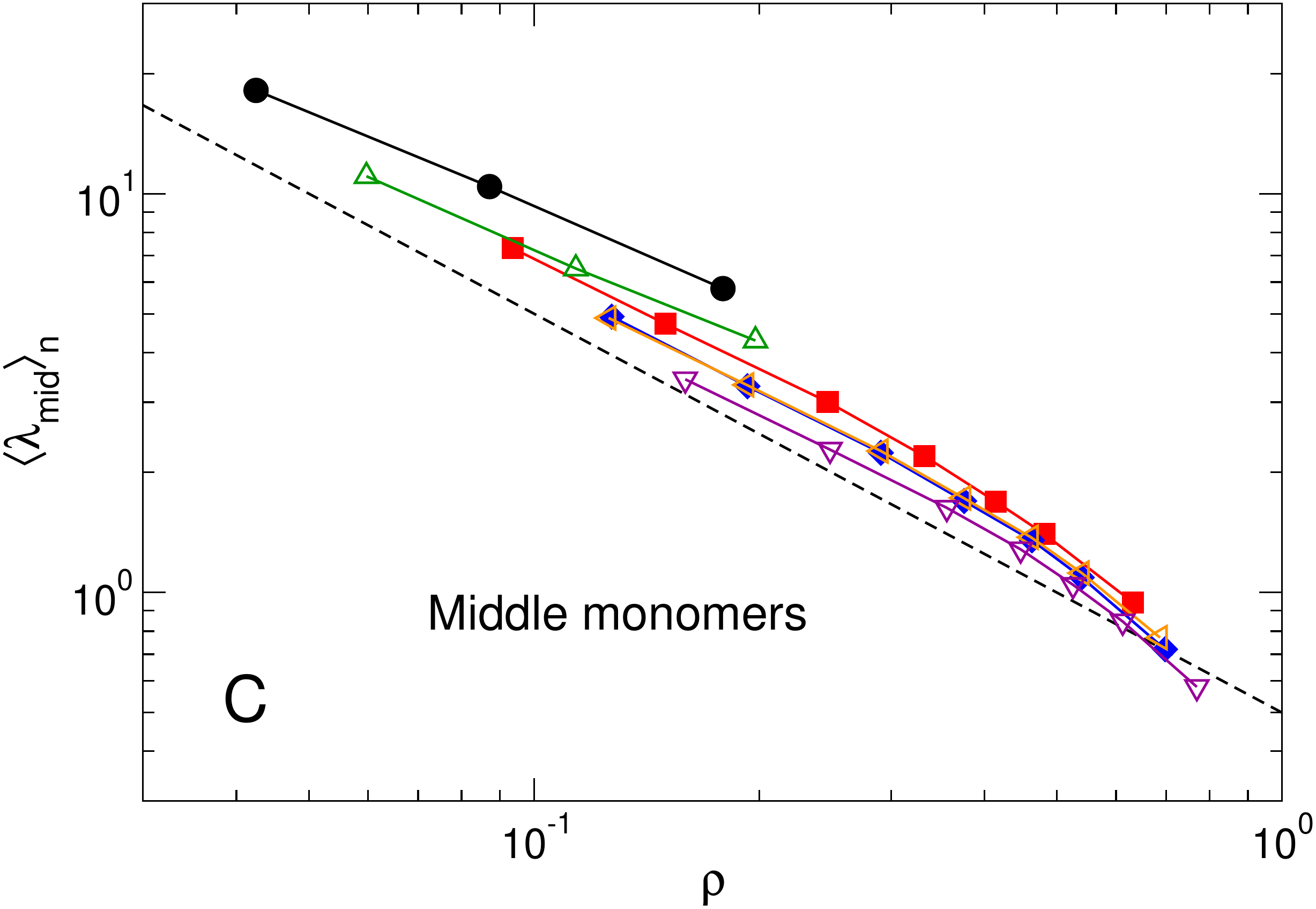}
\caption{Localization length of the bivalent particles  (A), the crosslinks (B), and the middle monomers of the odd-length strands (C). The dashed lines represent the function $(2 \rho)^{-1}$, and are reported to facilitate the comparison between the three panels.}
\label{fig:lambda_all}
\end{figure}

In Fig.~\ref{fig:lambda_all}, we report the localization length of the bivalent particles ($\lambda_2$, panel A) of the the crosslinks  ($\lambda_f$, panel B), and of the middle monomers of the odd-length strands ($\langle \lambda_\text{mid} \rangle_n$, panel C). We note that the total localization length $\lambda$ is given by $\lambda^2 = c \lambda_f^2 + (1-c) \lambda_2^2$, and that since $c$ is small in the simulated systems, we have $\lambda \simeq \lambda_2$, \textit{i.e.}, the total localization length is approximately equal to the localization length of the bifunctional particles.

\clearpage
\bibliography{bibliography.bib}
\end{document}